\documentclass[iop]{emulateapj}
\usepackage{amsmath}
\usepackage{amssymb}
\usepackage{mathbbol}
\usepackage{dsfont}
\usepackage{color}

\DeclareGraphicsExtensions{.eps,.eps.gz,.epsi}

\shorttitle{Age and mass of million stars from LAMOST}
\shortauthors{Xiang et al.}

\begin{document}

\title
{Ages and masses of million Galactic disk main sequence turn-off and sub-giant stars from the LAMOST Galactic spectroscopic surveys}
\author
{Maosheng Xiang\altaffilmark{1, 6},
 Xiaowei Liu\altaffilmark{2, 5},
 Jianrong Shi\altaffilmark{1, 4},
 Haibo Yuan\altaffilmark{3},
 Yang Huang\altaffilmark{2, 6},
 Bingqiu Chen\altaffilmark{2, 6},
 Chun Wang\altaffilmark{2},
 Zhijia Tian\altaffilmark{2, 6},
 Yaqian Wu\altaffilmark{3},
 Yong Yang\altaffilmark{2, 3},
 Huawei Zhang\altaffilmark{2},
 Zhiying Huo\altaffilmark{1},
 Juanjuan Ren\altaffilmark{1} 
}

\altaffiltext{1}{Key Laboratory of Optical Astronomy, National Astronomical Observatories, Chinese Academy of Sciences, Beijing 100012, P. R. China; email: msxiang@nao.cas.cn, sjr@bao.ac.cn}
\altaffiltext{2}{Department of Astronomy, Peking University, Beijing 100871, P. R. China; x.liu@pku.edu.cn}
\altaffiltext{3}{Department of Astronomy, Beijing Normal University, Beijing 100875, P. R. China}
\altaffiltext{4}{University of Chinese Academy of Sciences, Beijing 100871, P. R. China}
\altaffiltext{5}{Kavli Institute for Astronomy and Astrophysics, Peking University, Beijing 100871, P. R. China}
\altaffiltext{6}{LAMOST Fellow}

\journalinfo{submitted to The Astrophysical Journal Supplement}
\submitted{Received ; accepted }

\begin{abstract}{We present estimates of stellar age and mass for 0.93 million 
Galactic disk main sequence turn-off and sub-giant stars from the LAMOST Galactic 
Spectroscopic Surveys. The ages and masses are determined by matching with 
stellar isochrones using Bayesian algorithm, utilizing effective temperature 
$T_{\rm eff}$, absolute magnitude ${\rm M}_V$, metallicity [Fe/H] and $\alpha$-element 
to iron abundance ratio [$\alpha$/Fe] deduced from the LAMOST spectra. Extensive 
examinations suggest the age and mass estimates are robust. The overall sample 
stars have a median error of 34 per cent for the age estimates, and half of 
the stars older than 2\,Gyr have age uncertainties of only 20--30 per cent. 
Median error for the mass estimates of the whole sample stars is $\sim8$ per cent. 
The huge dataset demonstrates good correlations among stellar age, [Fe/H] ([$\alpha$/H]) 
and [$\alpha$/Fe]. Particularly, double sequence features are revealed in the 
both the age--[$\alpha$/Fe] and age--[Fe/H]([$\alpha$/H]) spaces. In the 
[Fe/H]--[$\alpha$/Fe] space, stars of 8--10\,Gyr exhibit both the thin and 
thick disk sequences, while younger (older) stars show only the thin (thick) 
disk sequence, indicating that the thin disk became prominent 8--10\,Gyr ago, 
while the thick disk formed earlier and almost quenched 8\,Gyr ago. 
Stellar ages exhibit positive vertical and negative radial gradients across 
the disk, and the outer disk of $R\gtrsim$\,9\,kpc exhibits a strong flare 
in stellar age distribution.}
\end{abstract}

\section{Introduction}

The Milky Way, as well as any other spiral galaxy, is an evolving system. 
Reliable age estimation for individual stars is therefore of great importance to secure 
a full understanding of the stellar population and assemblage history of the Galaxy. 
However, robust age estimates for large samples of Galactic field stars are 
still absent in spite of several large-scale surveys, both photometric and spectroscopic, 
having been carried out in the past decades, delivering positions, colors, spectral types, 
kinematics and chemistry for huge numbers of stars. The challenge is how to deliver 
realistic age estimates from those datasets, although huge in number, but often insufficient 
in accuracy. 

Stellar ages can hardly be `directly' measured but are generally inferred indirectly from 
photometric and spectroscopic observations in combination with stellar evolutionary 
models \citep[e.g.][]{Soderblom2010}. 
Asteroseismology has been demonstrated to be capable of delivering age estimates 
for individual stars with uncertainties at the level of about 10--20 per cent \citep[e.g.][]{ Gai2011, Chaplin2014}.
However, the method is only applicable to limited numbers of 
stars with sufficiently accurate, high cadence photometric measurements and to stars of 
a limited range of spectral types that exhibit prominent, solar-like oscillations. 
It has been suggested that carbon and nitrogen abundances can be 
age indicators for giant stars, but the reported results have been shown to have
large uncertainties, generally larger than 40 per cent \citep{Martig2016, Ho2016}. 
A practical way of robust age estimation for large samples of stars is via isochrone matching 
that match the observables with the predictions of stellar evolutionary models in the Hertzsprung-Russell (HR) diagram 
for given metallicity and elemental abundances. For this purpose, one needs accurate estimates
of atmospheric parameters, e.g. effective temperature $T_{\rm eff}$, surface gravity log\,$g$, 
absolute magnitude in $V$ band  ${\rm M}_V$ (or in other bands), metallicity [Fe/H] and 
alpha-element to iron abundance ratio [$\alpha$/Fe], derived for example from spectroscopy. 
The method works well mainly for main sequence turn-off (MSTO) or subgiant 
stars as for stars in those specific evolutionary stages, their atmospheric parameters vary 
significantly with age. The method is difficult for cool main-sequence or giant stars. 
Age estimates for stars in those evolutionary stages using this method could be dramatically 
wrong as isochrones of different ages are tightly crowded together. 

Limited by both observations and data analyses, for a long time robust age estimates via isochrone 
matching were only available for small samples of stars, of several hundred to a few thousand, in the 
solar neighborhood \citep[e.g.][]{Edvardsson1993, Nordstrom2004, Takeda2007, Haywood2013, Bergemann2014}. 
Only recently age estimates for hundreds of thousands of stars have been  
carried out utilizing the large stellar spectroscopic dataset from the LAMOST surveys \citep{Xiang+2015c}. 
\citet{Xiang+2015c} deduce ages for 300,000 MSTO stars spanning Galactic radii $7<R<15$\,kpc and 
heights $-3<Z<3$\,kpc, with a typical uncertainty of 30 per cent. The estimates of \citet{Xiang+2015c} 
were based on atmospheric parameters presented in the first release of value-added catalogs 
of the LAMOST Spectroscopic Survey of Galactic Anticentre \citep[LSS-GAC;][]{Liu+2014, Yuan+2015a} 
derived with the LAMOST Stellar Parameter Pipeline at Peking University \citep[LSP3;][]{Xiang+2015b}. 
The parameters, especially the log\,$g$ estimates suffer from significant uncertainties \citep{Ren+2016}, 
leading to some quite large age estimate errors \citep[e.g.][]{Wu2017}.     

In this work, we present age and mass estimates for nearly a million MSTO and subgiant stars. 
Compared to \citet{Xiang+2015c}, besides the significantly increased star 
number, the new estimates have benefited from several improvements: (1) The adopted 
basic stellar parameters deduced from LAMOST spectra are much more accurate thanks to 
dedicated efforts in improving both the spectral templates and the algorithms of the pipeline. 
In particular, values of $V$-band absolute magnitude ${\rm M}_V$ of individual stars are now 
directly delivered from the LAMOST spectra with machine learning method taking the 
LAMOST-{\em Hipparcos} common stars as a training data set, 
yielding ${\rm M}_V$ with uncertainties less than 0.3\,mag given good spectral quality \citep{Xiang+2017a, Xiang+2017b}. 
Estimates of [$\alpha$/Fe] abundance ratio from the LAMOST 
spectra have also become available; (2) A Bayesian approach is adopted to make use of 
priori knowledge of the stellar initial mass function for the age estimation, reducing bias 
of the estimated ages; (3) Extensive tests have been carried out to validate the age estimation, 
including a test with mock data, a comparison of the results with asteroseismic estimates and those inferred 
from the Gaia TGAS parallaxes, an examination using member stars of open clusters, and finally  
a robustness check using repeat observations. Note in this work we also provide robust 
mass estimates not available in \citet{Xiang+2015c}.  
Benefited from the huge sample and much improved parameter estimates, this work also explores 
the stellar age--[Fe/H]--[$\alpha$/Fe] correlations, as well as the variations of stellar age distribution 
across the Galactic disk. The sample will be publicly available via {\it http://lamost973.pku.edu.cn/site/data}.

This paper is organized as follows. Section\,2 introduces the LAMOST value-added catalogs, 
based on which our sample stars are defined. Section\,3 describes the selection criteria of  
MSTO and subgiant stars. The method of age and mass estimation is described in Section\,4. 
Examinations carried out to validate the age and mass estimates are presented in Section\,5. 
Section\,6 describes properties of the sample, including the distributions of 
stellar ages, masses and their errors, the age--[Fe/H]--[$\alpha$/Fe] correlations, as well as the spatial 
variations of age distributions across the disk. 
Section\,7 presents a discussion on how our sample could be affected by effects such as unresolved 
binaries and blue stragglers. Section\,8 is a brief summary.

\section{The LAMOST data}
\subsection{Value-added catalogs of the LAMOST Galactic surveys}
The LAMOST Galactic surveys \citep{Zhao+2012, Deng+2012} have several components focusing  
on different yet related aspects of Galactic studies, namely surveys of the LAMOST Galactic halo \citep{Deng+2012}, 
the Galactic Anticenter \citep[LSS-GAC;][]{Liu+2014}, stellar clusters \citep{Hou+2013}, 
and of the Kepler fields \citep{De_Cat+2015}. A survey of very bright stars utilizing 
grey and bright lunar conditions is also included. The raw 2D spectra collected for all the survey projects are processed 
uniformly with the LAMOST 2D reduction pipeline \citep{Luo+2015} to generate 1D spectra. 
Stellar parameters, including radial velocity $V_{\rm r}$, effective temperature $T_{\rm eff}$, 
surface gravity log\,$g$ and metallicity [Fe/H], are then derived from the 1D spectra with the LAMOST Stellar 
Parameter Pipeline \citep[LASP;][]{Wu+2014}. Both the 1D spectra and the LASP stellar parameters are
publicly available via the LAMOST official data releases \footnote{http://www.lamost.org} \citep{Luo+2012, Luo+2015}.

Since flux calibration by the default LAMOST 2D pipeline does not work well for plates of low Galactic latitudes, 
targeted by for example LSS-GAC spectra due to the unknown and significant extinction to the selected 
flux standard stars, an independent flux calibration pipeline has been developed at Peking University for LSS-GAC \citep{Xiang+2015a}. 
A stellar parameter pipeline, LSP3, has also been developed at Peking University that delivers, in addition to $V_{\rm r}$, $T_{\rm eff}$, log\,$g$ 
and [Fe/H] yielded by LASP, also values of [M/H], [$\alpha$/M], [$\alpha$/Fe], [C/H], [N/H], ${\rm M}_V$, 
${\rm M}_{K_{\rm s}}$, utilizing spectra processed with the LSS-GAC flux calibration pipeline \citep{Xiang+2015b, Liji+2016, Xiang+2017a}. 
Stellar parameters deduced with LSP3 for LSS-GAC targets, as well as values of extinction, distance 
and orbital parameters inferred using the LSP3 stellar parameters, are publicly released 
as LSS-GAC value-added catalogs\footnote{http://lamost973.pku.edu.cn/site/data} \citep{Yuan+2015a, Xiang+2017b}. 
Extensive examinations of stellar parameters yielded by LSP3 were carried out, and realistic parameter 
errors were assigned to each observation in a statistic way \citep{Xiang+2017a, Xiang+2017b}.    
For spectra of signal-to-noise ratios (SNRs) higher than 50, typical uncertainties of parameters of LSS-GAC DR2  
are about 5\,km/s for $V_{\rm r}$, 100\,K for $T_{\rm eff}$, 0.3\,mag for ${\rm M}_V$ and ${\rm M}_{K_{\rm s}}$, 
0.1\,dex for log\,$g$, [M/H], [Fe/H], [C/H] and [N/H], 0.05\,dex for [$\alpha$/M] and [$\alpha$/Fe], 0.04\,mag 
for $E(B-V)$ and 15 per cent for distance \citep{Xiang+2017b}.

\begin{figure}
\centering
\includegraphics[width=90mm]{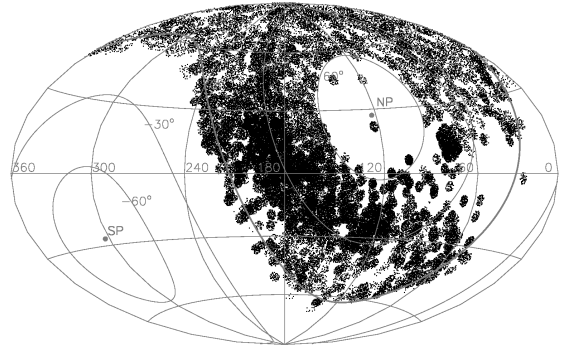}
\caption{Footprint of stars in the LAMOST value-added catalog. Only 1 in 50 stars are shown.}
\label{Fig1}
\end{figure}
Recently, we have applied the LSS-GAC flux calibration pipeline as well as Version 2 of the 
parameter determination pipeline LSP3 used to generate the LSS-GAC DR2 to all spectra 
of the LAMOST Galactic surveys collected by June, 2016. Basic stellar parameters, 
$V_{\rm r}$, $T_{\rm eff}$, $\log\,g$ and [Fe/H] yielded by the default LAMOST pipeline LASP, 
have been publicly released in December 2016 in the LAMOST DR4. 
Results from the LSP3 lead to a value-added catalog containing parameters 
derived from a total of 6.5 million stellar spectra of SNRs higher than 10, for 4.4 million unique stars. 
The database is used to define the MSTO and sub-giant star sample in the current work.
Fig.\,1 plots the spatial distribution of stars in this value-added catalog. 

\subsection{Choice of effective temperatures}
\begin{figure*}
\centering
\includegraphics[width=180mm]{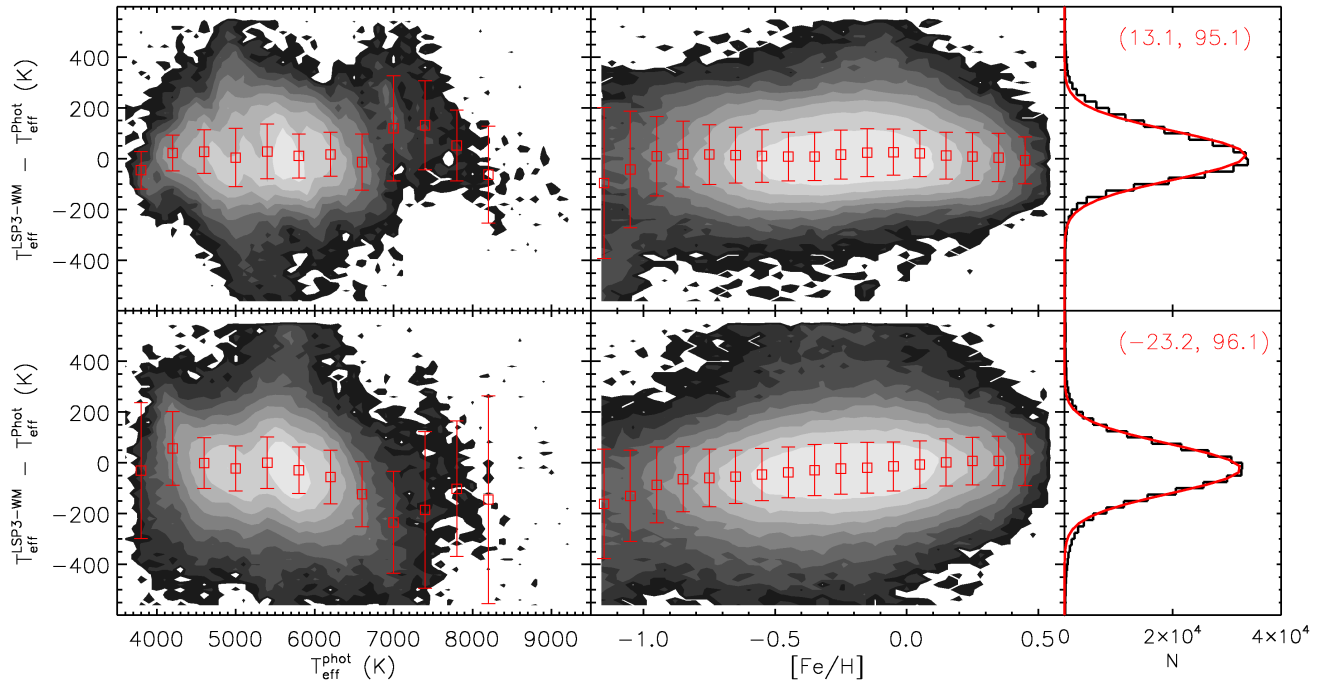}
\caption{Differences of LSP3 spectroscopic estimates of $T_{\rm eff}$  
and the photometric values deduced from the color-metallicity-temperature relations of \citet{Huang+2015b}, 
plotted against the photometric $T_{\rm eff}$ (left) and the spectroscopic (i.e. LSP3) [Fe/H] (middle). 
The upper panels show the LSP3 results based on the weighted-mean method, while the lower panels 
show those with the LSP3 KPCA method. 
Histogram of the differences, as well as Gaussian fits to the histograms are shown in the right 
panels. Means and 1-$\sigma$ dispersions of the Gaussians are marked in the plots.}
\label{Fig2}
\end{figure*}

\begin{figure*}
\centering
\includegraphics[width=180mm]{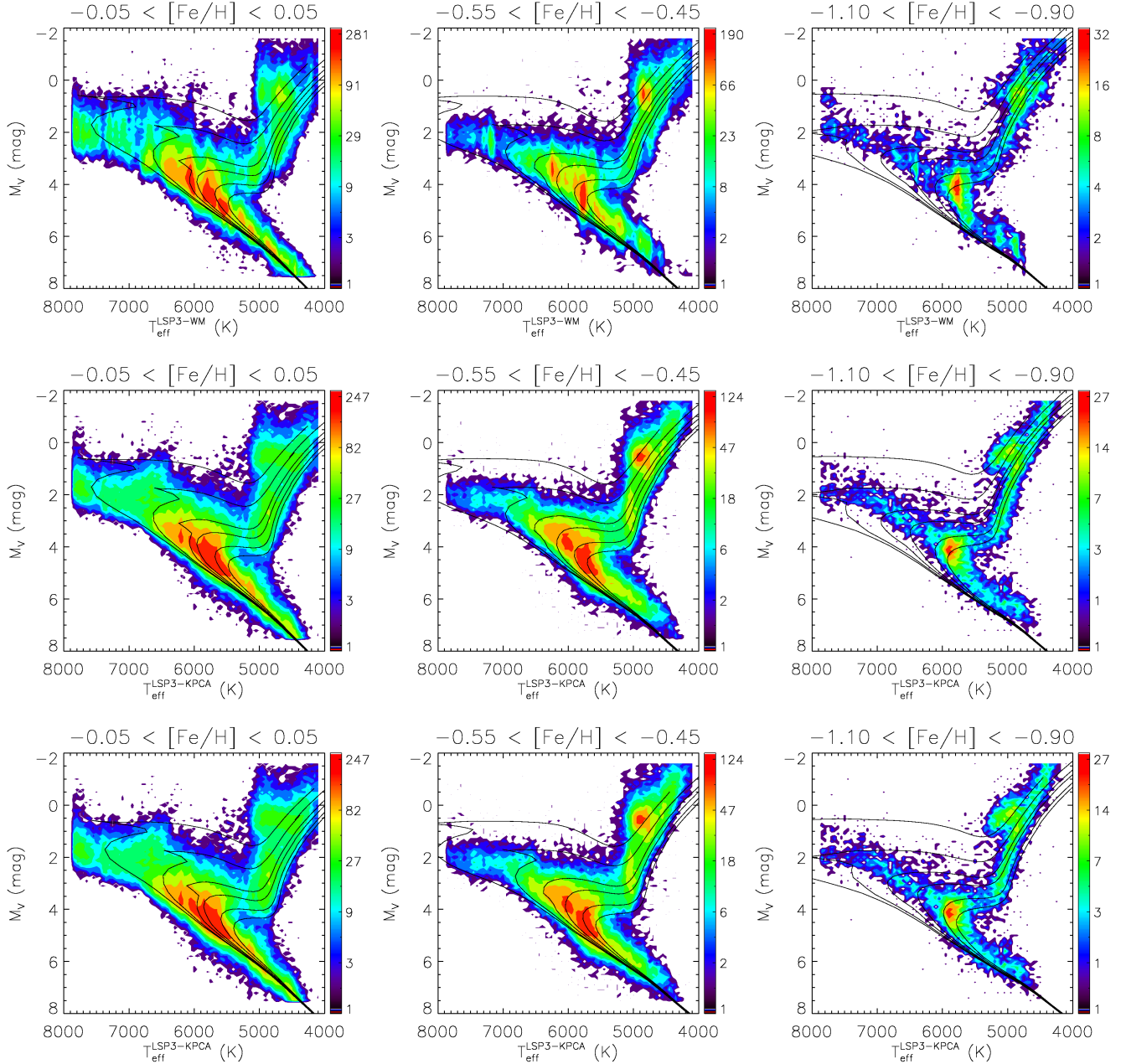}
\caption{Color-coded stellar number density distributions (in 10-based logarithmic scale) in the 
$T_{\rm eff}$--${\rm M}_V$ plane for stars in three [Fe/H] bins. Y$^2$ isochrones of ages 1, 2, 4, 6, 10 and 14\,Gyr 
are over-plotted. 
The top and middle panels show LSP3 estimates of $T_{\rm eff}$ with the weighted-mean and the KPCA 
method, respectively. The bottom panels show the same as the middle panels, except that the over-plotted isochrones 
have been calibrated to the temperature scale of \citet{Huang+2015b}.}
\label{Fig3}
\end{figure*}

\begin{figure}
\centering
\includegraphics[width=85mm]{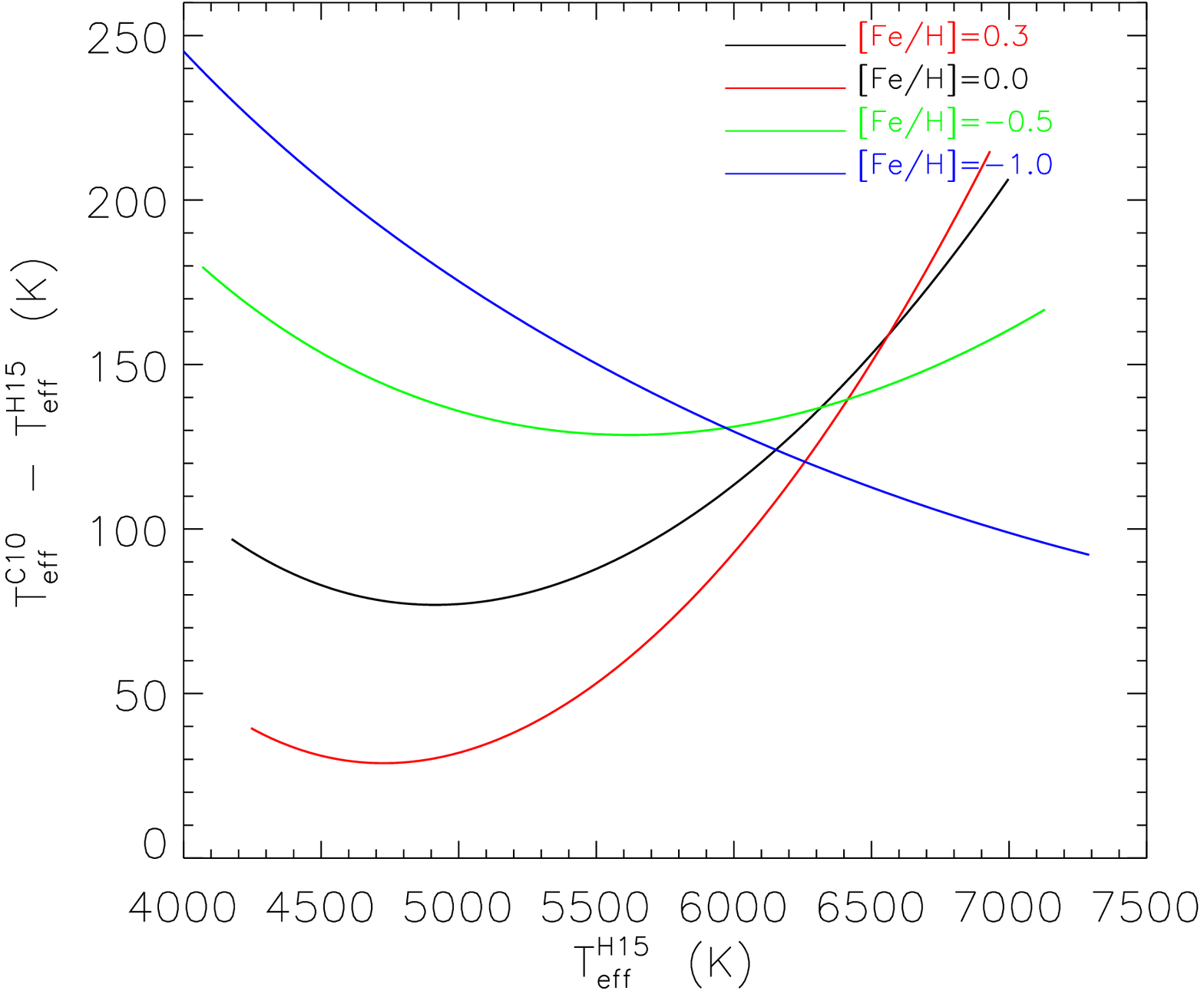}
\caption{Differences of temperature scales of \citet{Casagrande2010} and \citet{Huang+2015b} 
as a function of $T_{\rm eff}$ for different [Fe/H].}
\label{Fig4}
\end{figure}

Accurate estimates of $T_{\rm eff}$ are essential for age estimation, 
particularly for avoiding significant biases and systematic errors in the results. 
There are two sets of $T_{\rm eff}$ estimates in the value-added catalog, both estimated 
using the MILES empirical spectral template library \citep{Sanchez-Blazquez+2006} but 
with two different algorithms, the weighted-mean and the kernel-based principal component analysis (KPCA). 
Note that values of $T_{\rm eff}$ of the MILES template stars have been re-calibrated using 
the color-temperature-metallicity relation of \citet{Huang+2015b},
which is derived based on stars with direct, interferometric angular diameter measurements and {\em Hipparcos} parallaxes. 
The weighted-mean method works straight-forwardly with well controlled results, and yields robust $T_{\rm eff}$ 
estimates, even for a spectral SNR as low as 10 \citep{Xiang+2015b, Xiang+2017b}. 
On the other hand, due to the inhomogeneous distribution of the MILES stars in the parameter space, 
the weighted-mean estimates of $T_{\rm eff}$ suffer from the so-called clustering effects, 
at the level of a few tens to a hundred Kelvin. While estimates of $T_{\rm eff}$ yielded by the 
KPCA method do not suffer from significant clustering effects, the robustness of the results 
rely heavily on the SNRs of spectra under analysis \citep{Xiang+2017a}. 

Since no direct external calibration of $T_{\rm eff}$ has been carried out for LSS-GAC DR2, 
some further examinations of the $T_{\rm eff}$ estimates are desirable, especially considering 
that there could be some potential offsets in absolute scale between temperatures inferred from 
the color-temperature-metallicity relation of \citet{Huang+2015b} and those of the 
theoretical isochrones. For this purpose, here we 
first define a photometric sample from the value-added catalog. We select stars 
with photometric errors smaller than 0.03\,mag in SDSS $g$ and $r$-bands, and
with errors smaller than 0.04\,mag in 2MASS $K_{\rm s}$ band. We require the interstellar 
reddening $E(B-V)$ retrieved from the map of \citet[][hereafter SFD98]{Schlegel+1998} to have a value 
smaller than 0.05\,mag and that derived from  
the star-pair method in the value-added catalog to be smaller than 0.1\,mag, considering 
that for stars near the Galactic plane the SFD98 map sometimes yields too small, unrealistic values. 
We further require the stars to have a distance larger than 300\,pc as the SFD98 may have 
overestimated the reddening of stars of shorter distances. Stars observed with bad fibers of 
low throughputs or stars potentially affected by fiber crosstalk are discarded by requiring ${\rm BADFIBER}=0$ 
and ${\rm SNR-BRIGHTSNR}>-150$. Here `BRIGHTSNR' is the highest SNR 
of stars targeted by the 5 fibers adjacent to the fiber of the star of concern. 
Finally, we select only stars with a spectral SNR higher than 20. 
The above criteria lead to a total of 350,000 stars that form our photometric sample. 

Fig.\,2 plots the differences of LSP3 estimates of $T_{\rm eff}$ and photometric values  
inferred from $g-K_{\rm s}$ colors using the relation of \citet{Huang+2015b},  
as a function of $T_{\rm eff}^{\rm phot}$ or [Fe/H]. Here the $g-K_{\rm s}$ colors are 
dereddened using SFD98 values of $E(B-V)$ and the extinction coefficients of \citet{Yuan+2013}. 
The figure shows that for the temperature range 3800--6300\,K, $T_{\rm eff}$ values 
yielded by both the weighted-mean and the KPCA methods are in good agreement with the photometric values, 
with differences $<50$\,K. At higher temperatures, the KPCA method yields values that 
are smaller than the photometric ones by 100--200\,K, while the weighted-mean method 
yields results that are higher than the photometric values by 150\,K 
for stars around 7000--7500\,K. No systematic trends of differences with [Fe/H] are seen 
for $T_{\rm eff}$ yielded by the weighted-mean method in the [Fe/H] range $-1.0$ -- 0.5\,dex, 
the applicable metallicity range of the color-metallicity-temperature relation of \citet{Huang+2015b}. 
However, for $T_{\rm eff}$ estimated with the KPCA method, there is a positive trend of  
difference with [Fe/H]. Nevertheless, the dispersions of the differences 
amount to only $\sim$95\,K for both sets of LSP3 temperature estimates. 
Considering that a bias in $T_{\rm eff}$ estimates, especially a positive trend of 
$T_{\rm eff}$ with [Fe/H] may lead to some undesired systematics in the age estimates 
and in the distribution of stars in the age--metallicity space, we choose to correct for the 
biases, albeit small, in the KPCA $T_{\rm eff}$ estimates. The correction is done in the 
$T_{\rm eff}$--[Fe/H] plane by interpolating a grid of bias values created with the photometric sample. 

Fig.\,3 plots the logarithmic (base 10) number density of stars in the $T_{\rm eff}$ -- ${\rm M}_V$ 
diagrams for three metallicity bins. To better illustrate the potential systematic patterns, 
only stars with spectral SNRs higher than 50 are shown.  
The Yonsei-Yale \citep[Y$^2$;][]{Demarque2004} 
isochrones are overplotted. The figure shows that for stars of solar metallicity bin, 
the distribution in the $T_{\rm eff}$ -- ${\rm M}_V$ diagram is basically consistent with isochrones. 
In contrast, for stars in the two metal-poor bins, the distributions deviate from the theoretical isochrones.
For example, most (88 per cent) stars of ${\rm M}_V>4.0$\,mag
in the $-1.0<{\rm [Fe/H]}<-0.9$\,dex bin have temperatures of about 200\,K lower than 
values of the isochrones of 14\,Gyr, older than the dynamic age of the universe \citep[13.8\,Gyr, e.g.][]{Planck2016}. 
The offsets are significant and can not be caused by random errors of the stellar parameters only. 
The apparent [Fe/H]-dependent inconsistencies are undesired and could have severe impacts 
on our sample selection, age estimates and subsequently statistical analysis.  
Similar deviates are seen when the isochrones of the Dartmouth Stellar Evolution 
Database \citep[DESP;][]{Dotter+2008} or the PAdova and TRieste Stellar Evolution Code 
\citep[PARSEC;][]{Bressan2012} are used. We suspect that the offsets are caused by 
different temperature scales of the color-temperature-metallicity relation of \citet{Huang+2015b}
and the theoretical isochrones. The color-temperature-metallicity relation of \citet{Huang+2015b} 
is based on `directly' measured temperatures, while the values of $T_{\rm eff}$ of theoretical isochrones 
depend on the stellar atmospheric models adopted. 
The figure also shows that, as mentioned above, values of $T_{\rm eff}$ yielded by the weighted-mean 
method suffer moderate clustering effects. In the current work, we thus adopt  
the KPCA estimates of $T_{\rm eff}$ in order to avoid any potential patterns in the age estimates due to 
the clustering effect in the temperature estimates.

It is of particular interest to unravel the causes of the [Fe/H]-dependent discrepancies 
of temperature scales between the direct measurements and the theoretical isochrones, 
as it may help understand the robustness of those currently widely used stellar 
atmospheric models, especially those of metal-poor stars. In fact, regardless of [Fe/H], 
\citet{Huang+2015b} have found an overall systematic difference of about 100\,K 
between the temperatures given by their photometric relations and those derived from methods 
based on the stellar atmospheric models in literature \citep[e.g.][]{Santos2004, Valenti2005, Casagrande2010}.  
This important issue is however out of the scope of this paper. As a temporary remedy 
to avoid potential biases in our age estimation, here we have
opted to adjust the temperature scale of isochrones to match that of \citet{Huang+2015b}. 
In doing so, we have implicitly assumed that the isochrone temperature scale is the same as that given by  
\citet{Casagrande2010} utilizing the infrared flux method (IRFM), which also relies on stellar atmospheric models. 
Fig.\,4 plots the differences of IRFM $T_{\rm eff}$ of \citet{Casagrande2010} 
and those of \citet{Huang+2015b} for different metallicities. 
It is obvious that the differences depend on temperature and metallicity. 
At solar metallicity, the IRFM scale of \citet{Casagrande2010} gives $T_{\rm eff}$ that is $\sim$100\,K 
higher than values yielded by the relation of \citet{Huang+2015b} for a temperature of $\sim$5800\,K. The difference 
is consistent with the finding of \citet{Huang+2015b}. However, the trends of differences 
with metallicity and temperature shown in the figure cannot be ignored for robust and unbiased age 
estimation. We correct the isochrone temperatures to match the scale of \citet{Huang+2015b} for 
each metallicity ${\rm [Fe/H]}>-1.2$\,dex, but leave the more metal-poor isochrones 
untouched. The disposition will not yield any inconsistency in our results because in 
the current work we are concerned only with disk stars of ${\rm [Fe/H]}>-1.0$\,dex, and 
most of our sample stars have [Fe/H] higher than $-0.8$\,dex. Nevertheless, we have 
carefully examined the age estimation for stars of low metallicities. A test shows that if we instead 
calibrate the LSP3 spectroscopic temperatures to match the IRFM scale and estimate the ages 
using the isochrones without temperature corrections, the resultant ages do not deviate from 
the current estimates by any significant amounts ($<1$\,Gyr). 
Note that the IRFM temperature scale of \citet{Casagrande2010} is only applicable to a limited 
color range ($0.78< V-K_{\rm s} < 3.15$). To overcome the limitation, the isochrone grids of 
hotter temperatures are corrected for using the amount of temperature corrections at the boundary. 
This simplification again will not cause any significant impact in our results as hot ($T_{\rm eff}>7000$\,K) 
stars are young ($\lesssim$1\,Gyr) and a 100--200\,K difference in temperature will cause 
only very small changes in the age estimates. 
Finally, we have to point out that the IRFM temperature scale of \citet{Casagrande2010} 
is only appropriate for dwarf and subgiant stars but not for giant stars. 
Since we focus our work on MSTO and subgiant stars, this limitation does not affect the current work. 

The bottom panels of Fig.\,3 compare the data and the isochrones 
after the temperature corrections. The plots show much better agreement, not only for the 
metal-poor bins, but also for the solar-metallicity bin. Most of the MSTO stars are now 
encompassed by 14\,Gyr isochrones. 
Nevertheless, for the metal-poor, low-temperature ($T_{\rm eff}\lesssim5300$\,K) main sequence 
stars, there are still some discrepancies between the data and the isochrones. Those offsets 
could either be due to possible different temperature scales of the isochrones and the IRFM 
calibration of \citet{Casagrande2010}, or caused by overestimates of isochrone absolute 
magnitudes for those metal-poor, low-temperature stars. However, these remaining discrepancies 
are not expected to have any significant impact on our results because we focus on MSTO and subgiant 
stars, and our target selection criteria ($\S{3}$) have effectively excluded those cool, 
metal-poor main sequence stars.

\section{sample selection}
\begin{figure}
\centering
\includegraphics[width=85mm]{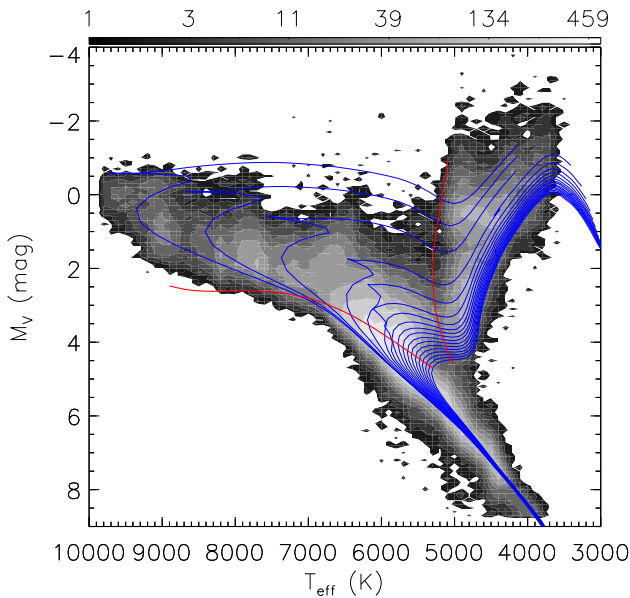}
\caption{Example of sample selection criteria in the $T_{\rm eff}$--${\rm M}_V$ diagram 
for the solar metallicity bin. 
The background grey scale map shows the number density (in 10-based logarithmic scale) of stars of $-0.05<{\rm [Fe/H]}<0.05$\,dex 
in the value-added catalog. Blue curves, from left to right, are Y$^2$ isochrones of ages ranging from 1 to 16\,Gyr 
for [Fe/H]=0.0 and [$\alpha$/Fe] = 0.0\,dex. Stars falling within the two red curves constitute the main sequence 
turn-off and subgiant sample stars for this solar metallicity bin.}
\label{Fig5}
\end{figure}
We define a sample of MSTO and subgiant stars for which reliable stellar ages can be  
determined via isochrone fitting. In doing so, we first trace the locus of MSTO 
in the $T_{\rm eff}$ -- ${\rm M}_V$ plane using the Yonsei-Yale (Y$^2$) isochrones, 
similar to the technique of \citet{Xiang+2015c} except in the latter case the analysis was 
carried out in the $T_{\rm eff}$ -- log\,$g$ plane instead. 
This is done for isochrones with [Fe/H] from $-3.0$\,dex to $+0.5$\,dex with a step of 0.1\,dex.  
For each [Fe/H], an [$\alpha$/Fe] value is adopted, which increases from 0.0\,dex for ${\rm [Fe/H]} \ge 0.0$\,dex  
to 0.3\,dex for ${\rm [Fe/H]}\leq-1.0$\,dex. 
Similarly, the trajectory of the base of red giant branch (RGB) is also determined 
in the $T_{\rm eff}$ -- ${\rm M}_V$ plane. 
Let ${\rm M}_V^{\rm TO}$, a function of $T_{\rm eff}$, denote the trajectory 
of the MSTO, and $T_{\rm eff}^{\rm bRGB}$, a function of ${\rm M}_V$, 
denote the trajectory of the base of RGB. The sample stars are then defined by requiring, 
\begin{equation}  
T_{\rm eff} > T_{\rm eff}^{\rm bRGB} + \Delta{T_{\rm eff}},
\end{equation}
\begin{equation}  
{\rm M}_V < {\rm M}_V^{\rm TO} + \Delta{\rm M}_V,
\end{equation}
where  $\Delta{T_{\rm eff}}$ is set to reduce contamination from RGB stars 
due to the errors in $T_{\rm eff}$ estimation, and is set to be a constant of 300\,K. 
$\Delta{\rm M}_V$ is set to be a function of $T_{\rm eff}$: $\Delta{\rm M}_V = 0.0005\times(T_{\rm eff} - T_{\rm eff}^{\rm MINISO})$, 
where $T_{\rm eff}^{\rm MINISO}$ is the minimum temperature of MSTO of isochrones 
for a given set of [Fe/H] and [$\alpha$/Fe]. Fig.\,5 plots an example of the criteria for [Fe/H] = 0 and [$\alpha$/Fe] = 0\,dex. 
Our choice of $\Delta{\rm M}_V$ ensures that main sequence stars of high temperature 
(e.g. $> 6500$\,K) are also included in our sample, as their ages can be well-estimated. 
Note that trajectories of the MSTO and base RGB in the $T_{\rm eff}$ -- ${\rm M}_V$ plane, 
as well as the adopted $T_{\rm eff}^{\rm MINISO}$, for different metallicities are listed in the Appendix.

To select sample stars from the value-added catalog, we first discard stars of saturated spectra,  
by requiring ${\rm SATFLAG}=0$, stars potentially suffering from significant fiber cross-talking,   
by requiring ${\rm SNR - BRIGHTSNR} > -150$, and stars observed with bad fibers, by 
requiring ${\rm BADFIBER}=0$. For stars with duplicate observations, only results based on 
the spectrum of highest SNR are selected. Results of unique stars are then grouped into [Fe/H] 
bins of width 0.1\,dex. MSTO and subgiant sample stars of the individual metallicity bins 
are then selected using the criteria defined by Eqs.\,(1) and (2). 
To reduce potential contamination from giant and supergiant stars, as well as to ensure 
the robustness of stellar parameters used to defined the sample, we require that the sample 
stars must have $T_{\rm eff} < 10000$\,K, log\,$g > 3.0$\,dex, ${\rm [Fe/H]}>-1.0$\,dex 
and a spectral SNR higher than 20.  
Note that several sets of stellar parameters are provided in the value-added catalog.
The parameters adopted here refer to the recommended ones. 
The [Fe/H] cut above is used to discard metal-poor halo stars, whose LSP3 stellar parameters 
may need some further improvement, and thus to leave us with a pure disk star sample.  
Finally, for each metallicity bin, stars that stray into the area in the $T_{\rm eff}$ -- ${\rm M}_V$ 
plane beyond the boundary defined by isochrones of age 16\,Gyr towards the direction of 
lower temperatures, are discarded.  
With the above criteria, a total of 932,313 unique MSTO and subgiant sample stars are selected. 
Here and later, we use the term `MSTO-SG stars' to denote 
those selected MSTO  and subgiant stars for convenience. 
We note that there are 420,000 duplicate observations of these MSTO-SG stars in the value-added catalog. 

\begin{figure}
\centering
\includegraphics[width=85mm]{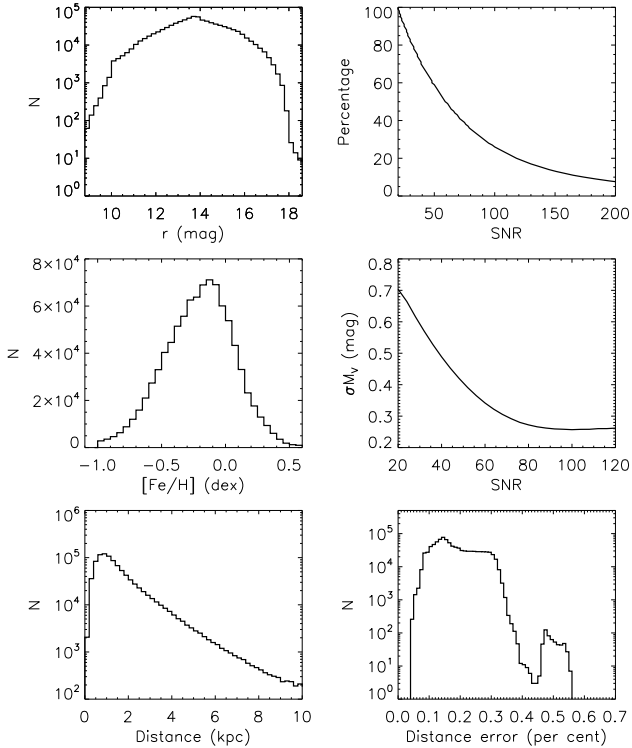}
\caption{From upper-left to bottom right are respectively the distribution of $r$-band magnitudes, 
the cumulative percentage distribution of spectral SNRs, the distribution of [Fe/H], errors of ${\rm M}_V$ 
estimates as a function of SNR, the distributions of distances and distance errors, of the MSTO-SG 
star sample.}
\label{Fig6}
\end{figure}
Fig.\,6 plots the distributions of $r$-band magnitudes, SNRs, [Fe/H], ${\rm M}_V$, 
distances and distance errors of the MSTO-SG sample stars. The stars 
have $r$-band magnitudes ranging from 9 to 18\,mag, peaking at $\sim$14\,mag. 
This is owning to the large number of very bright stars of $r<14$\,mag observed 
utilizing the grey/bright lunar nights. 
The sample stars cover a wide range of spectral SNRs, and about 60 per cent of the stars 
have a SNR higher than 50. 
The [Fe/H] distribution peaks at about $-0.1$\,dex, and less than 2 per cent of the stars 
are more metal-poor than $-0.8$\,dex. 
Errors of ${\rm M}_V$ estimates are sensitive to the SNR, with typical values of 0.7\,mag 
for a SNR of 20, and decrease to 0.25\,mag at SNRs$\gtrsim80$. The median value of 
${\rm M}_V$ errors of the whole sample stars is 0.37\,mag. 
About half of the stars have a distance smaller than 1.2\,kpc, and about a quarter 
more distant than 2\,kpc. The fraction of stars more distant than 3\,kpc is 11 per cent. 
The median value of the relative distance errors is 16.8 per cent, with 38 per cent of 
the stars having a relative distance error smaller than 15 per cent, and 21 per cent  
having a relative distance error larger than 25 per cent.

\section{Method}
Stellar age and mass are estimated via matching the observed stellar parameters with 
theoretical isochrones with a Bayesian scheme similar to that of \citet{Lindegren2005}. 
The stellar parameters include $T_{\rm eff}$, ${\rm M}_V$, [Fe/H] and [$\alpha$/Fe]. 
Note that although the value-added catalog also provides estimates of 
${\rm M}_{K_{\rm s}}$, here we have opted not to use both ${\rm M}_V$ and ${\rm M}_{K_{\rm s}}$ 
for the age estimation because $M_{\rm V}$ and ${\rm M}_{K_{\rm s}}$ are found to be 
largely correlated with each other. 
Similarly, log\,$g$ is not used because it carries largely the 
same information as $M_{\rm V}$ considering that the log\,$g$ values of template/training stars 
used by LSP3 for log\,$g$ estimation are constrained mainly by the {\em Hipparcos} parallaxes, 
i.e., the same as those used for the estimation of absolute magnitudes. 
Including ${\rm M}_{K_{\rm s}}$ and log\,$g$ in the  
age estimation might in principle help constrain the results, but this works only  
if the above correlations can be dealt with properly, and we intend to leave this to a future work. 
The value-added catalog provides also two sets of [$\alpha$/Fe] estimates 
for MSTO-SG stars, one ([$\alpha$/Fe]$_1$) utilizing spectra of 3910--3980, 4400--4600 and 5000--5300\,{\AA}, 
the other ([$\alpha$/Fe]$_2$) utilizing 4400--4600 and 5000--5300\,{\AA} only \citep[cf.][]{Xiang+2017b}. 
Considering that the inclusion of 3910--3980{\AA} may cause artifacts for low temperature 
stars as the Ca\,{\sc ii}\,H, K lines are usually saturated, and also that the [$\alpha$/Fe]$_1$ are 
found to exhibit larger random errors for high temperatures ($\gtrsim6000$\,K) stars, we choose 
to use [$\alpha$/Fe]$_2$ throughout the paper if not specified.     

\subsection{Stellar isochrones}
We have chosen to use the Yonsei-Yale (${\rm Y}^2$) isochrones \citep[V2;][]{Demarque2004} 
for the age estimation. The ${\rm Y}^2$ isochrones cover a wide range of stellar ages (0.001--20\,Gyr), 
which is convenient for us to apply a Bayesian algorithm as biases induced by the abrupt age cutoff 
of the isochrone grids could be negligible.
Moreover, the ${\rm Y}^2$ isochrones have grids of different [$\alpha$/Fe], allowing 
us to make use of the [$\alpha$/Fe] measurements to better constrain the ages. 
Grids of the ${\rm Y}^2$ isochrones are interpolated into a uniform set of grids of step 
0.1\,dex in [Fe/H] and [$\alpha$/Fe], 0.1\,Gyr 
in age for ${\rm age} < 1$\,Gyr, 0.2\,Gyr for $1 < {\rm age} < 2$\,Gyr and 0.5\,Gyr 
for ${\rm age} > 2$\,Gyr, utilizing the interpolator provided by \citet{Demarque2004}. 
The isochrones adopt the color table of \citet{Lejeune1998} and assign colors and magnitudes 
in $UBVRIJHKLL'M$ photometric bands for each grid model. Here the $UBV$ system is that of \citet{Buser1978}, 
of which the $V$-band agrees well with the Johnson \citep[e.g.][]{Bessell2005}, 
$RI$ that of \citet{Bessell1979}, and $JHKLL'M$ that of \citet{Bessell1988}. 

There are also quite a few other sets of stellar isochrones that are widely used,
such as the Dartmouth Stellar Evolutionary Database \citep[DSEP;][]{Dotter+2008} 
and the PAdova and TRieste Stellar Evolution Code \citep[PARSEC;][]{Bressan2012}. 
Different isochrones are based on more or less 
different stellar model assumptions, and thus may lead to some different age estimates.
Generally, the different isochrones yield stellar ages for MSTO stars with some systematic differences
of the order of about 1\,Gyr, along with dispersions at the same level \citep[e.g.][]{Haywood2013, Xiang+2015c, Hills2015}.

\subsection{Age and mass estimation} 
The observed properties of a star are largely determined by three parameters,  
namely age ($\tau$), initial stellar mass ($M$) and chemical compositions ($Z$). 
In Bayesian theory, their joint (posterior) probability density function thus can be written as,  
\begin{equation}
  f(\tau,M, Z) = Af_0(\tau,M, Z)L(\tau, M, Z),
\end{equation}
where $f_0$ is the priori density distribution, $L$ is the likelihood function, and $A$ is 
a normalization factor to ensure $\int\int\int f(\tau, M, Z){\rm d}\tau{\rm d}M{\rm d}Z = 1$.

Let $\mathbf{O}$ denote the observed stellar parameters {$T_{\rm eff}$, ${\rm M}_V$, [Fe/H], [$\alpha$/Fe]}, 
and $\mathbf{P}$ denote the isochrone values given $\tau$, $M$ and $Z$. 
The likelihood function $L$ is then given by, 
\begin{equation}
L(\tau, M, Z) = \prod_{i=1}^n \frac{1}{\sqrt{2\pi}\sigma_i} \times \exp\left(-\frac{\chi^2}{2}\right), 
\end{equation}
\begin{equation}
\chi^2 = \sum_{i=1}^n \left(\frac{O_i-P_i(\tau, M, Z)}{\sigma_i}\right)^2, 
\end{equation}
where $n$ is the number of observables, and 
$\sigma_i$ is the Gaussian error of the $i$-th observed parameter.

For the priori density distribution, we adopt the same formula as used by \citet{Lindegren2005}, 
\begin{equation}
f_0(\tau, M, Z) = \psi(\tau)\phi(Z\mid\tau)\xi(m\mid Z,\tau).
\end{equation}
Here $\psi(\tau)$ is the star formation history, $\phi(Z\mid\tau)$ is the metallicity distribution 
as a function of age, and $\xi(m\mid Z, \tau)$ is the initial mass function (IMF) as a function of 
metallicity and age. In principle, the distributions of $\psi(\tau)$ and $\phi(Z\mid\tau)$ 
should be a function of position across the Milky Way. Considering that the star formation rate 
and the metallicity distribution as a function of age as well as of spatial position, are not well known, 
we have adopted a flat distribution for both $\phi(Z\mid\tau)$ and $\xi(m\mid Z, \tau)$ to avoid 
potentially large biases in the resultant age estimates. The IMF is better 
known -- the star formation process yields more low mass stars than massive stars and the number 
of stars as a function of mass can be generally well described by power-laws or log-normal \citep[e.g.][]{Salpeter1955, 
Kroupa2001, Chabrier2003}. Here we have adopted the IMF form of \citet{Kroupa2001},
\begin{equation}
\xi(m)\propto m^{-a},
\end{equation} 
where $a = 0.3$ for $m < 0.08M_{\odot}$,  $a = 1.3$ for $0.08 < m < 0.5M_{\odot}$ and $a = 2.3$ 
for $m > 0.5M_{\odot}$. We assume that the IMF is invariant with age and metallicity. 
For Galactic field stars, this may not be a bad assumption \citep[e.g.][]{Kroupa2001, Kroupa2013}, especially 
considering that our sample stars cover a rather limited mass range.

For each age, the joint probability is then evaluated using the isochrones grids with parameter values 
within $\pm$3$\sigma$ of the observed ones, 
and the age of the star of concern is then estimated by taking the mean of the distribution with error 
given by the standard deviation. An alternative age estimate is obtained by taking the 
mode of the joint probability distribution. In the latter case, the error of age is estimated by requiring 
that the 1$\sigma$ error covers 68\% of the area of the joint probability distribution. 
In some cases where the parameters are poorly estimated, the resultant joint probability has a broad distribution, 
peaking either near the young or old age cutoffs of the isochrones. 
As a consequence, the resultant mean ages tend to fall by the middle of the age interval of the isochrones, 
while the mode ages tend to have a value close to the upper or lower boundary of the age interval.   
A comparison of the two age estimates helps one to evaluate the quality of age estimation. 
Such cases do not occur often, as most of our sample stars have well determined parameters 
that fall within the suitable range of age estimation. 
The mean ages thus derived are analyzed in the Sections below. 
The mass estimate is also taken as the weighted-mean value given by all isochrone grids within $\pm$3$\sigma$ 
of the observed stellar parameters. 
Here the weights for mass estimates are the same as those for the age estimates. 
Note that for both age and mass estimation, effects from unevenly spaced age grids have been 
considered via multiplying the joint probability by $\Delta\tau$, age space of the grids.

[$\alpha$/Fe] of the Y$^2$ isochrone grids is limited in the range of 0.0--0.6\,dex. 
As many stars have [$\alpha$/Fe] values close to or smaller than 0.0\,dex, an [$\alpha$/Fe] cutoff 
of the isochrones causes a cutoff in the joint probability distribution function, 
which induce bias in the age and mass estimates.  
To avoid such bias, we opt not to use [$\alpha$/Fe] when calculating the 
joint probability. Instead, we calculate ages for isochrones of [$\alpha$/Fe] values of 0.0, 0.2 and 0.4\,dex 
separately, and then estimate the final age by linearly extrapolating (interpolating for stars with [$\alpha$/Fe] 
between 0 and 0.4\,dex) the results to match the observed [$\alpha$/Fe].

\section{validation of age and mass estimates}
\subsection{Test with mock data}
 \begin{figure}
\centering
\includegraphics[width=90mm]{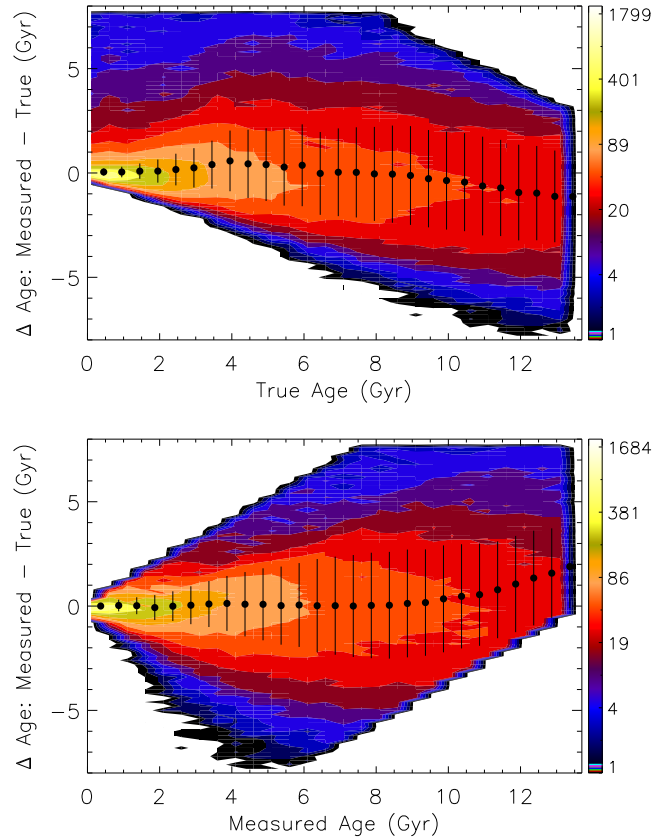}
\caption{Differences of estimated and true ages as a function of 
the latter (upper) and former (lower) for the mock sample. The color-coded contours 
show stellar number densities in logarithmic (base 10) scale. Mean and standard deviations 
of differences for the individual age bins are shown by dots with error bars.}
\label{Fig7}
\end{figure}
As an examination of the method, we estimate stellar age and mass for 
a mock data set generated with Monte-Carlo simulation, and compare the results 
with true values. To generate the mock data, 35 sets of isochrones are first selected, 
each with a given combination of {$\tau$, [Fe/H], [$\alpha$/Fe]}. The isochrones cover 
an age range of 0.2--13.5\,Gyr, with a step of 0.2 and 0.5\,Gyr respectively 
for isochrones of ages below and above 1\,Gyr. The isochrones cover an 
[Fe/H] range of $-1.8$--0.3\,dex and an [$\alpha$/Fe] range of 0.0--0.4\,dex, 
with the older, more metal-poor isochrones having higher [$\alpha$/Fe] values. 
For each set of isochrones, mock stars for a total mass of 50000 $M_{\odot}$ are 
retrieved following the IMF of \citet{Kroupa2001}. 
Gaussian errors are added to the retrieved $T_{\rm eff}$, ${\rm M}_V$ and 
[Fe/H], with dispersions of 130\,K, 0.4\,mag, and 0.15\,dex, respectively. 
Note that these values of dispersions correspond to typical, not minimum, 
errors of the MSTO-SG sample stars.
 
\begin{figure}
\centering
\includegraphics[width=90mm]{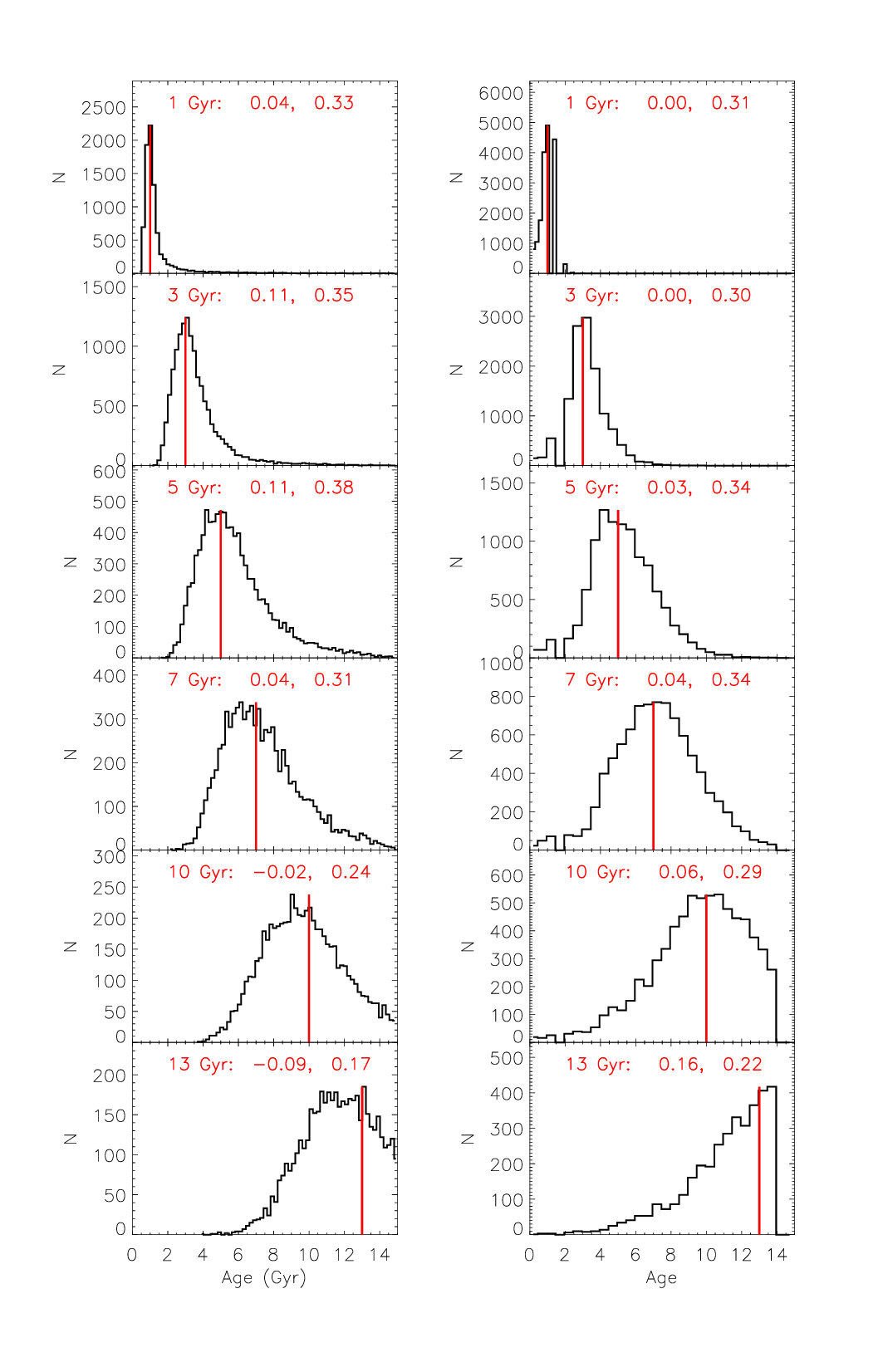}
\caption{$Left$: Distributions of estimated ages for stars of given true ages marked 
by the vertical red lines in the individual panels. Values of the true age, as well as the 
percentage difference and standard deviation are marked. 
$Right$: Same as left, but for distributions of true ages for stars of estimated ages in a 
limited ($\pm0.25$) range.}
\label{Fig8}
\end{figure}

\begin{figure}
\centering
\includegraphics[width=90mm]{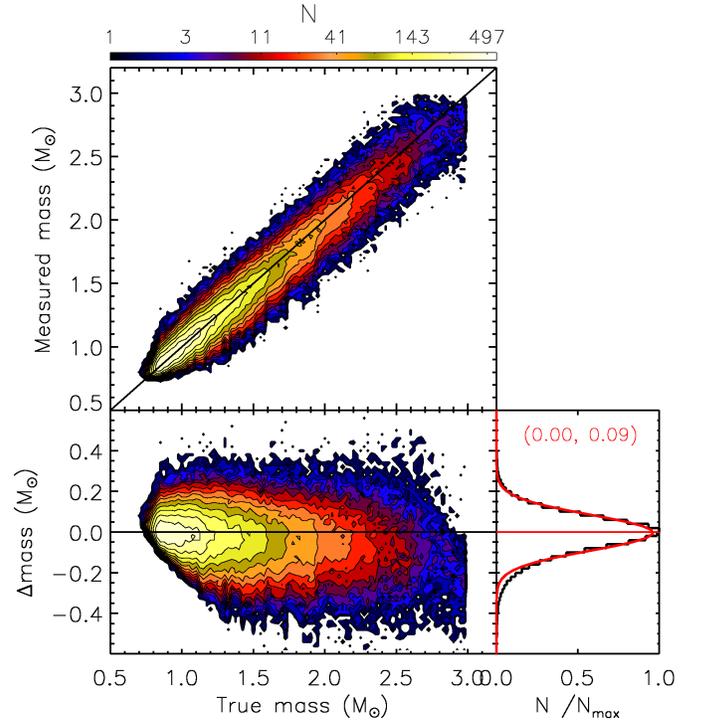}
\caption{Comparison of measured and true masses for the mock data set. 
The mean and dispersion of a Gaussian fit to the differences are marked in 
a plot on the right.}
\label{Fig9}
\end{figure}

MSTO and sub-giant stars are selected from the mock catalog, and their ages and masses 
are estimated with the Bayesian method described above. Fig.\,7 plots the differences of 
measured and true ages as a function of either the former or the latter. The figure shows 
that the mean differences are close to zero at all ages except for the oldest stars. 
Ages of truely oldest ($>12$\,Gyr) stars are systematically underestimated by $\sim$1\,Gyr, 
while stars of oldest measured ages are in fact generally 1--2\,Gyr younger. 
The standard deviations of the differences increase from $\sim$0.6\,Gyr 
at age 2\,Gyr to 2.3\,Gyr at age 8\,Gyr, then flatten.  
There are a small fraction of young stars whose ages are significantly overestimated 
and a small fraction of old stars whose ages are significantly underestimated due to their large 
parameter errors. As a result, for stars of measured ages
between approximately 4 and 9\,Gyr, their true ages may spread over a wide range, 
although the number of stars having large age errors is expected to be small. 
Fig.\,8 plots one-dimensional distribution of the measured ages for stars 
with the same true age, as well as distribution of the true ages for stars within 
a given range of measured age. For a given true age, the distribution of measured ages 
is clearly non-Gaussian but exhibits a tail at the older end, a consequence of the uneven 
distribution of isochrones in the $T_{\rm eff}$ -- ${\rm M}_V$ plane. 
Stars of the oldest measured ages show a tail of small values in the distribution of their true ages, 
mainly caused by the cutoff of true age of isochrones at 13.5\,Gyr. Typical percentage values of the  
mean differences are a few per cent, with typical standard deviations of 25 per cent 
for old stars and 35 per cent for young stars. 

Fig.\,9 plots a comparison of estimated and true masses. The figure shows 
very good consistency, with small systematic differences ($<0.05$\,$M_{\odot}$) for 
mass range 0.7--3.0\,$M_{\odot}$, along with standard deviations 
only $\sim$\,0.09\,$M_{\odot}$, indicating typical relative mass errors smaller than 10 per cent.

Similar analyses were carried out for other sets of parameter errors corresponding 
to MSTO-SG sample stars having different spectral SNRs. For example, for parameter error set 
of 150\,K, 0.5\,mag, 0.15\,dex for respectively $T_{\rm eff}$, ${\rm M}_V$ and [Fe/H], 
the results show that the standard deviations of age differences between the measured and 
true values are only slightly ($\lesssim$5 per cent) increased compared to the above quoted values. 
For the parameter error set of 100\,K, 0.3\,mag, 0.1\,dex, 
which correspond to spectral SNRs$\gtrsim$60, 
the standard deviations are only 20 per cent for old stars, and 25 per cent for 
young stars. This is encouraging since more than one third of the sample stars 
have parameter errors smaller than these values. 

\subsection{Comparing age and mass estimates with seismic values}
 \begin{figure}
\centering
\includegraphics[width=85mm]{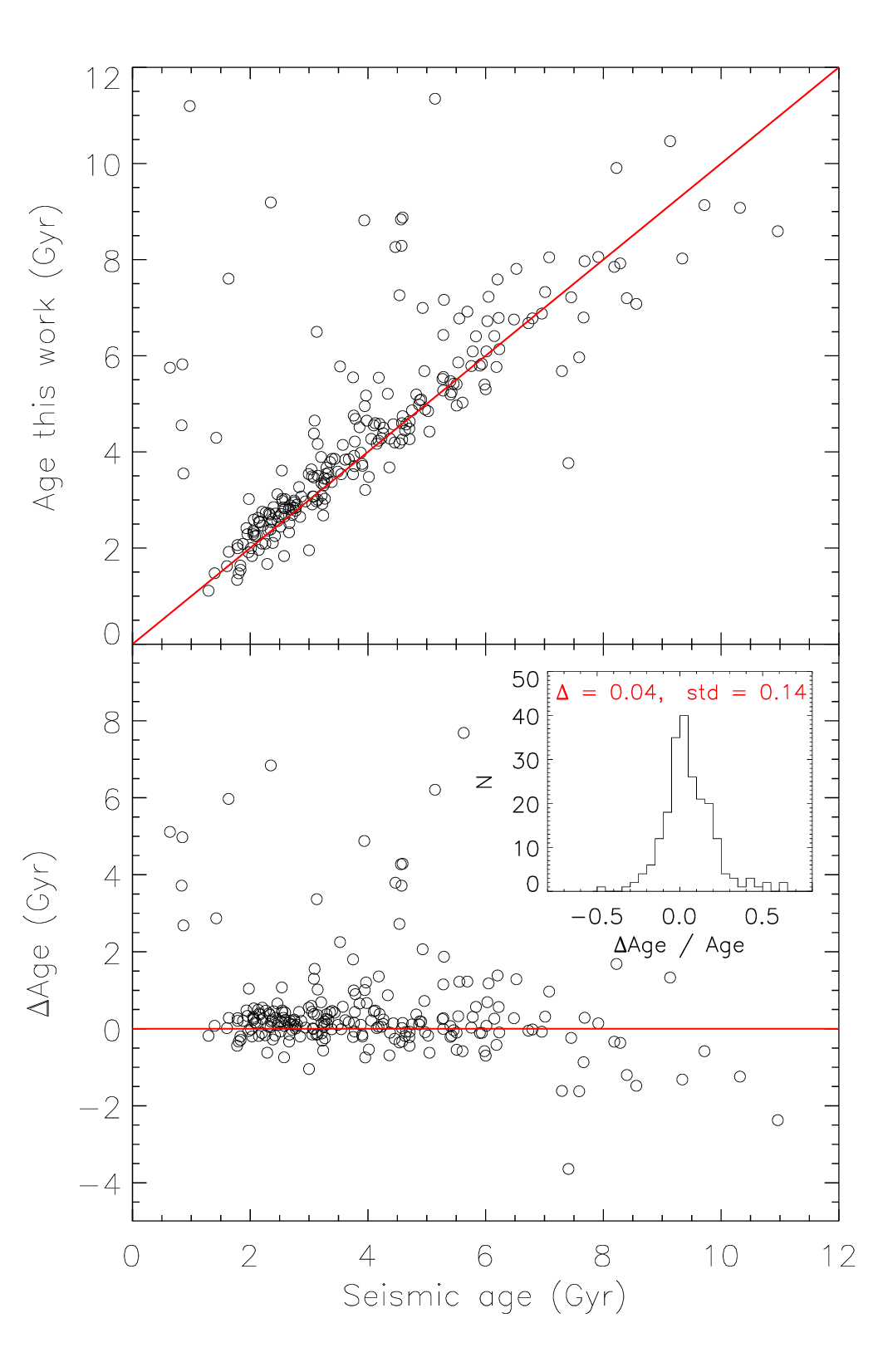}
\caption{Comparison of ages estimated using the approach of the current work and those 
derived based on the asteroseismic log\,$g$.}
\label{Fig10}
\end{figure}

\begin{figure}
\centering
\includegraphics[width=85mm]{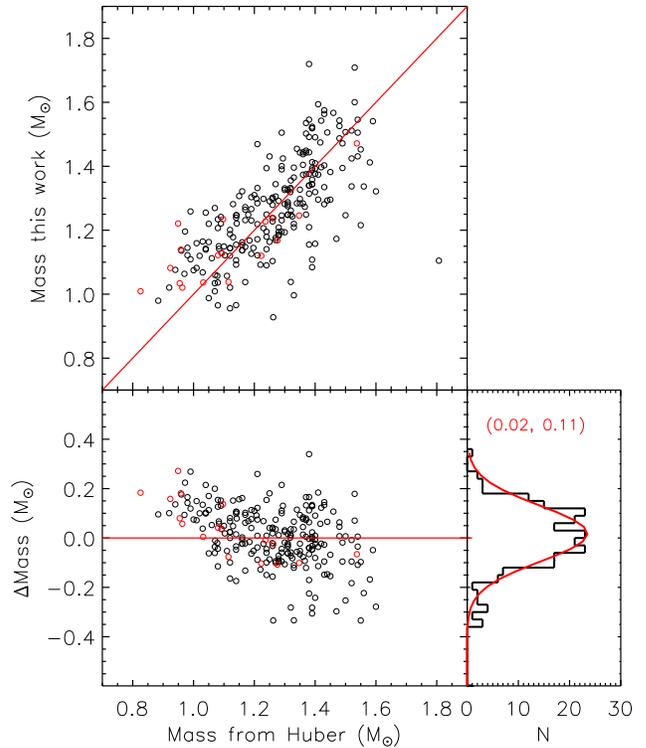}
\caption{$Left$: Comparison of mass estimates in the current work 
and those of \citet[][black circles]{Huber+2014} 
and those deduced from the asteroseismic scaling relation (red circles).
A Gaussian fit to the differences 
yields a mean and dispersion of $0.02$ and 0.11\,$M_{\odot}$, respectively. }
\label{Fig11}
\end{figure}

Stellar asteroseismology is suggested to yield log\,$g$ with uncertainties smaller 
than 0.05\,dex \citep[e.g.][]{Creevey2013, Hekker2013, Chaplin2014}. 
We thus expect that stellar ages derived from asteroseismic log\,$g$ are more 
accurate than -- and therefore can be used to test -- the current estimates. 
For our sample, 230 stars have asteroseismic log\,$g$ 
measurements available from the catalog compiled by \citet{Huber+2014}. 
For those stars, we have determined their ages using the asteroseismic log\,$g$ 
measurements along with our spectroscopic estimates of $T_{\rm eff}$, [Fe/H] and [$\alpha$/Fe]. 
Fig.\,10 compares ages estimated using ${\rm M}_V$ (`Age of this work') with those 
based on the asteroseismic log\,$g$ measurements. 
The figure shows good consistency between the two sets of age estimates. 
The mean value of percentage differences for the whole sample is quite small, 
along with a standard deviation that is only 14 per cent. Nevertheless, for old stars, ages 
estimated using the approach of the current work tend to be underestimated by $\sim$1\,Gyr compared to values 
derived with the asteroseismic log\,$g$, consistent with results shown in Fig.\,7 for the mock data set.
There is also a small fraction of stars for which the current age estimates are significantly larger than 
the asteroseismic log\,$g$ based estimates. 
Some of these stars are likely subgiant or red clump stars whose absolute 
magnitudes and effective temperatures from the LAMOST spectra are over-estimated.
Note that for both sets of age estimates, the same set of parameters $T_{\rm eff}$, [Fe/H] 
and [$\alpha$/Fe] are used, so the differences seen in Fig.\,10 reflect the errors induced by the uncertainties in ${\rm M}_V$ only.  

Fig.\,11 compares our mass estimates with those of \citet{Huber+2014} 
for the 230 common stars, as well as with stellar masses directly inferred from the asteroseismic 
scaling relation for a subset sample of 17 stars whose scaling relation based masses 
have a propagated random error smaller than 0.2\,$M_{\odot}$. 
$T_{\rm eff}$ adopted in this work is used for inferring masses from the scaling relation. 
Masses of Huber et al. are derived using asteroseismic log\,$g$, photometric $T_{\rm eff}$ 
and [Fe/H] utilizing multiple evolutionary tracks, but mainly the DESP tracks. Interestingly, 
although different sets of stellar parameters and isochrones have been used in deriving those 
stellar masses, the figure shows quite good agreement, with a mean difference of 
only 0.02\,$M_{\odot}$ and a dispersion of 0.11\,$M_{\odot}$. Nevertheless, at the 
high mass end, there is a subset of stars for which our current mass estimates are systematically 
lower by 0.25\,$M_{\odot}$. Those stars are also the outliers found in Fig.\,10. 
At the low mass end, it seems that the current estimates yield masses $\sim$0.1--0.2\,$M_{\odot}$ 
larger. Similar result is also seen in the comparison with masses derived from  
the scaling relation. For those stars, duplicate observations of LAMOST, as well as an 
examination of ${\rm M}_V$ values derived using parallaxes from the Tycho-Gaia astrometric solution
\citep[TGAS;][]{Michalik2015, Lindegren2016}, suggest that our current estimates are robust. 
We thus suspect that the discrepancies are probably caused by random errors in
asteroseismology based mass estimates or by systematic errors in either the 
asteroseismic parameters or the asteroseismic scaling relation at the low mass end.
In fact, for those low mass stars, masses inferred from scaling relation have a 
typical propagated error of $\sim$0.2$M_{\odot}$, which is significantly larger 
than the errors of our current mass estimates ($\sim$0.05\,$M_{\odot}$).  

\subsection{Comparison with ages derived from the Tycho-Gaia parallaxes}
 \begin{figure}
\centering
\includegraphics[width=85mm]{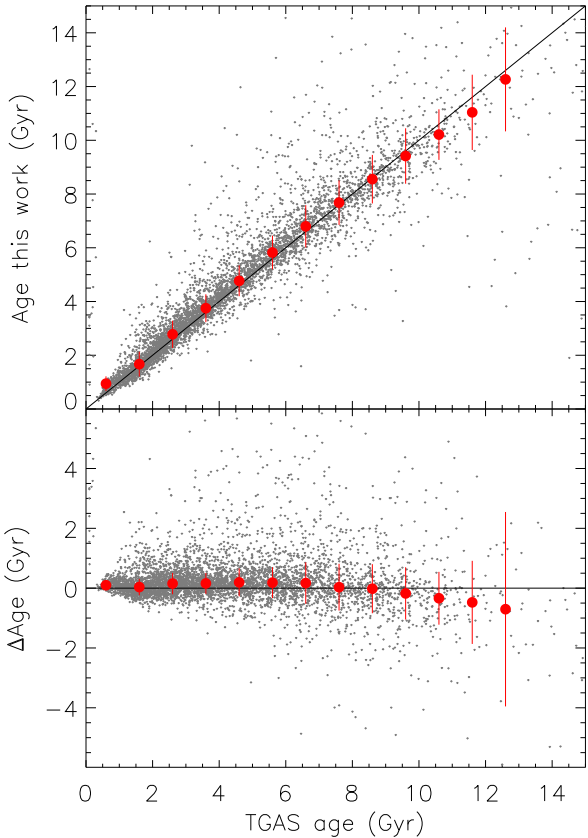}
\caption{Comparison of age estimates of the current work and those derived 
utilizing absolute magnitudes inferred from the TGAS parallaxes.}
\label{Fig12}
\end{figure}
Accurate parallaxes from the Gaia TGAS catalog \citep{Brown2016, Lindegren2016} provide 
independent determinations of absolute magnitudes and thus can be used to test our age estimates. 
A cross-identification of our value-added catalog with the TGAS catalog yields more than 
0.3 million common stars, and about 50,000 of them have values of ${\rm M}_V$ inferred from 
the TGAS parallaxes with errors smaller than 0.2\,mag. 
For those common stars, \citet{Xiang+2017b} have compared values of ${\rm M}_V$ 
and distances with those derived utilizing the TGAS parallaxes, and found very good agreement. 
Here we further derive stellar ages using ${\rm M}_V$ inferred from the 
TGAS parallaxes and parameters $T_{\rm eff}$, [Fe/H] and [$\alpha$/Fe] 
from the value-added catalog to test the robustness of our age estimates. 

After applying an error cut of 0.2\,mag for ${\rm M}_V$ inferred from the TGAS parallaxes, 5258 unique stars  
in common with the TGAS catalog remain in our MSTO-SG star sample. Fig.\,12 compares our age estimates 
for these MSTO-SG stars with ages estimated utilizing ${\rm M}_V$ inferred from the TGAS parallaxes.  
The figure shows good agreement, with a mean value of  
percentage differences close to zero and a standard deviation 
of only 14 per cent. Similar to the results shown in Fig.\,7 for the mock data, there is a small fraction of stars 
for which our results seem to be significantly overestimated, probably due to the large uncertainties of their 
atmospheric parameter and absolute magnitude estimates. Note that for both sets of age estimates, 
the same values of atmospheric parameters $T_{\rm eff}$, 
[Fe/H] and [$\alpha$/Fe] are used. So any discrepancies revealed by the comparison 
are likely mainly caused by the uncertainties in ${\rm M}_V$ estimates only.

\subsection{Test with open clusters}
 \begin{figure}
\centering
\includegraphics[width=80mm]{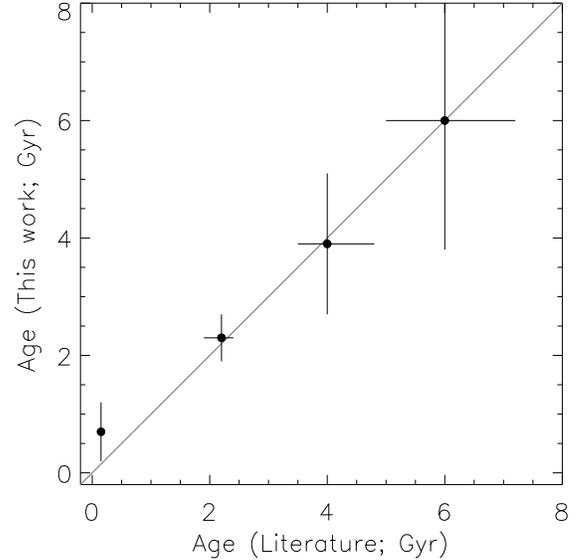}
\caption{Comparison of age estimates of open clusters with literature values. 
The vertical error bars represent dispersions (standard deviations) of age estimates of the individual 
cluster member stars, while the horizontal error bars represent age estimates in literature.}
\label{Fig13}
\end{figure}

\begin{table*}
\centering
\caption{Age estimates of open clusters.}
\label{}
\begin{tabular}{lllllllrll}
\hline
Cluster       & ${\rm Age}_{\rm Liter}$  &  $\Delta {\rm Age}_{\rm Liter}^a$ & Age  & $\sigma$(Age)    & [Fe/H] & $\sigma$[Fe/H] & $(m-M)_0$ & $E(B-V)$ & Number of stars$^{b}$        \\
       &           (Gyr)         &  (Gyr) &      (Gyr)        & (Gyr)    &  \\
\hline
M35  &  0.15       & 0.1 -- 0.2 &   0.7 & 0.5                & $-0.14$  & 0.19 & 9.75 & 0.35      &  395 (217)  \\   
NGC2420 & 2.2     &  1.9 -- 2.4&  2.3 & 0.4  & $-0.31$ & 0.09& 11.70 & 0.04 & 34 (24) \\
M67              &    4.0      & 3.5 -- 4.8  &  3.9 & 1.2          &$-0.04$ & 0.08 & 9.80 & 0.03       &  982 (184) \\
Berkeley32    &  6.0        & 5.0 -- 7.2              &   6.0 & 2.2       & $-0.44$ & 0.10 & 12.44 &0.21 &  18 (17) \\
\hline
\end{tabular}
\begin{description}
\item[$^a$] $References$
     M35: \citet{von_Hippel2002, Kalirai2003a, Meibom2009};  \\
     NGC2420: \citet{Demarque1994,Twarog1999};   \\
     M67: \citet{Demarque1992, Carraro1994, Dinescu1995, Fan1996, Richer1998, 
VandenBerg04, Schiavon2004, Balaguer2007, Sarajedini2009, Barnes2016}; \\
     Berkeley32: \citet{Kaluzny1991, Richtler2001, Salaris2004, DOrazi2006, Tosi2007}. \\
\item[$^b$]  Including duplicate observations, and number of unique stars is shown in the brackets.  \\
\end{description}
\end{table*}

Open clusters in the Milky Way are generally believed to form from a monolithic gas cloud 
on short timescales, so that member stars of a cluster belong to a single-aged population.
Ages of cluster members thus provide an independent test of the robustness of our age estimation. 
For this purpose, a number of LAMOST plates have been designed to target open clusters of different 
ages utilizing grey nights reserved for monitoring the instrument performance. Together  
with data from the main surveys, we are able to select MSTO-SG stars in four 
open clusters, namely M35 (NGC2168), NGC2420, M67 (NGC2682) and Berkeley32. 
These clusters cover an age range from $\sim$100\,Myr to 6\,Gyr. 

A detailed description of member star identification for these open clusters will be presented 
elsewhere (Yang et al. in preparation). Brieflly, for M67 and Berkeley32, member stars 
are identified by combining LAMOST radial velocities and UCAC4 proper motions \citep{Zacharias+2013}. 
For M35 and NGC2420, contaminations from background stars are so 
severe that kinematics alone is insufficient for robust member identification, so additional 
constrains from the distance moduli are used to discard background stars that deviate significantly 
($>1.5$\,mag) from the peak values of the clusters.  
The numbers of member stars that pass our selection criteria of MSTO-SG stars   
are listed in Table\,1. 
Note that here we have excluded some blue stragglers that also pass our selection criteria 
of MSTO-SG stars but whose ages may have been significantly underestimated (cf. Section\,7). 
The measured ages, metallicities, extinction values and distance moduli of the clusters, 
obtained by taking means of the individual member stars, as well as age estimates in the literature  
are also listed in Table\,1. 

Fig.\,13 presents a direct comparison between the measured cluster ages and the literature values. 
The figure shows that our age estimates are largely in good agreement with the literature values, mostly 
derived by isochrone fitting of color-magnitude diagrams. 
The relatively large dispersion of ages of the individual member stars of Berkeley\,32 is mainly 
due to the small number of member stars identified in this cluster. For the young open cluster M\,35, 
the relatively large deviation of our estimate from the literature values is caused by 
net overestimates of stellar ages for individual `MSTO' stars at low temperatures, 
as our sample selection criteria at low temperatures prefer both stars whose ${\rm M}_V$ 
are underestimated due to random errors and background MSTO stars with old ages.

\subsection{Comparison of results from duplicate observations}
\begin{figure}
\centering
\includegraphics[width=85mm]{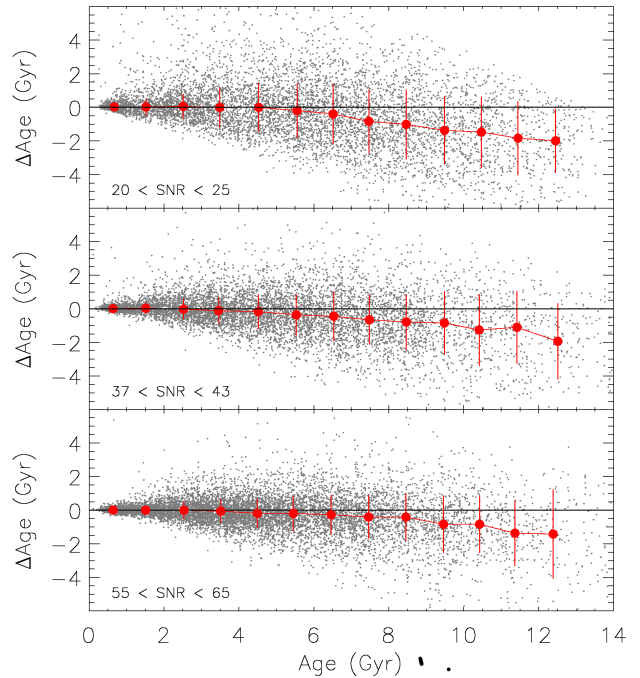}
\caption{Differences of ages deduced from the duplicate and default observations. 
Different panels show results for stars in different SNR bins, as marked in the plots. 
Red dots and error bars show median values and standard deviations of the 
differences in the individual age bins.}
\label{Fig14}
\end{figure}

\begin{figure}
\centering
\includegraphics[width=85mm]{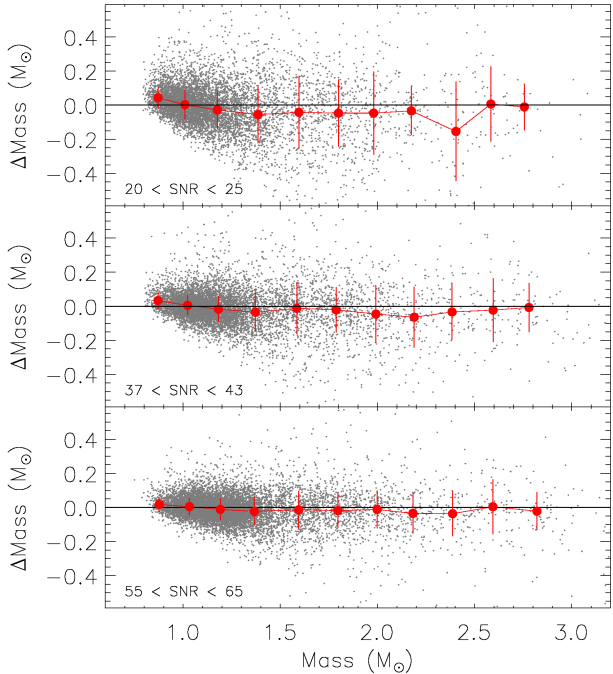}
\caption{Same as Fig.14, but for mass estimates.}
\label{Fig15}
\end{figure}
As mentioned in \S{3}, there are more than 400,000 duplicate observations for the MSTO-SG 
sample stars. As a sanity check, Fig.\,14 plots the differences of age estimates between  
those duplicate observations and the sample stars. 
Only duplicate observations that have SNRs comparable (within 20 per cent) to those of the sample stars 
are used in the comparison, and the comparisons 
are carried out for different SNR bins. 
The figure shows that for young (e.g. $<6$\,Gyr) stars, the duplicate observations yield 
ages in excellent agreement with those deduced from the default observations. For old stars, 
duplicate observations yield ages 1--2\,Gyr younger, depending on the SNRs. 
This deviation is due to the combination of the effects of the uneven distribution of 
the isochrones in the $T_{\rm eff}$--${\rm M}_V$ plane and errors in ${\rm M}_V$ 
estimates. As a result of the uneven distribution of the isochrones in the $T_{\rm eff}$--${\rm M}_V$ plane, 
deviation of the estimated ${\rm M}_V$ from the true value of an `exact' MSTO star 
for a given effective temperature always yield underestimated age. 
The dispersions of the differences are small, generally less than 20 per cent at all 
ages. Note that both the default observations that define the sample and the duplicate observations  
contribute to the dispersions, and thus the random errors of ages estimated from the 
default observations as induced by the spectral noises are expected to be smaller than the dispersions (by a factor of $\sim1.4$).

Fig.\,15 plots the differences of masses estimated from the default and the duplicate observations. 
The agreement is quite good. Typical dispersions are a few to ten per cent, depending on the SNRs.

\section{Properties of the sample}

\subsection{Distributions of ages and masses}
\begin{figure}
\centering
\includegraphics[width=85mm]{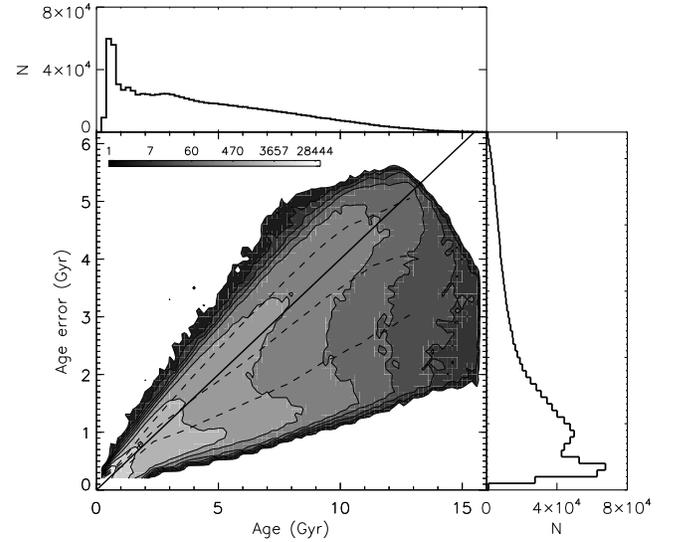}
\caption{Distribution of ages and errors of the sample stars. 
The contours show stellar number densities in logarithmic (base 10) scale. The solid line 
delineates 40 per cent age errors, and the dashed lines, from left to right, show median errors as a function 
of age for stars within spectral SNR bins 20--25, 37.5--42.5 and 55--65, respectively.}
\label{Fig16}
\end{figure}

\begin{figure}
\centering
\includegraphics[width=90mm]{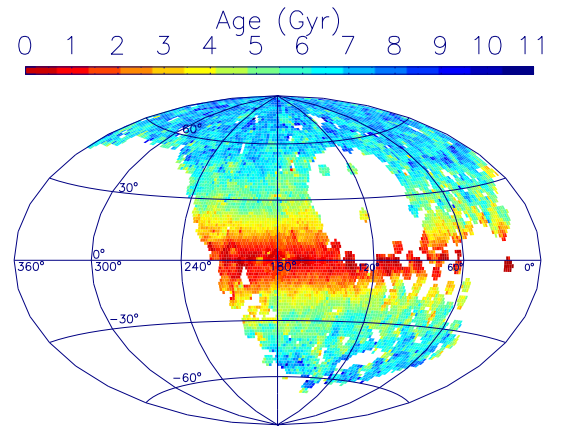}
\caption{Distribution of median stellar ages in Galactic coordinate system ($l$, $b$). 
The data are divided into patches of 1.5$^\circ$$\times$1.5$^\circ$ to draw the map.}
\label{Fig17}
\end{figure}

\begin{figure}
\centering
\includegraphics[width=85mm]{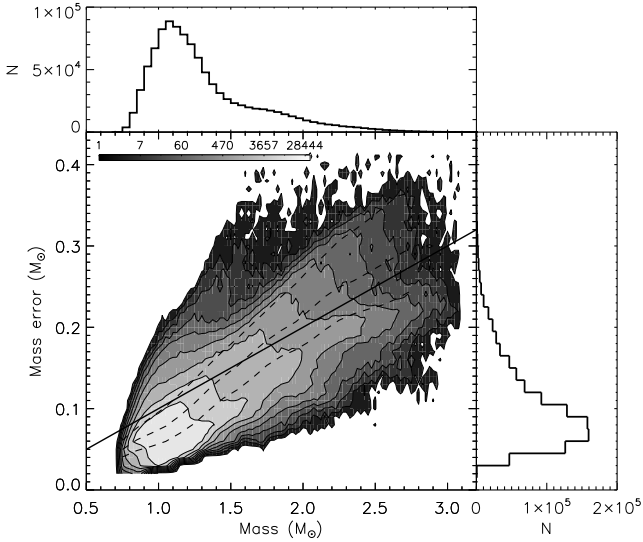}
\caption{Same as Fig.\,16 but for mass estimates. The solid line delineates 10 per cent mass errors.}
\label{Fig18}
\end{figure}
Fig.\,16 plots the distributions of estimated ages and errors of the sample stars. The 
age distribution shows that there are more young stars than old ones in our sample. In particular, there is 
a peak of stars younger than 1\,Gyr. Note that the distribution is a consequence of both the age distribution 
of the underlying stellar population and the selection effects of the observations as well as the sample definition, 
rather than simply the former. Both the observations and the SNR (at 4650\AA) cut prefer young 
stars as they are bright and blue, and thus tend to have high SNRs. Our selection criteria used to define sample 
stars in the HR diagram also prefer young stars. A detailed analysis of the selection effects 
of the sample is quite complicated and out of the scope of the current paper. The errors of age estimates 
vary sensitively with the SNR. For stars of SNRs higher than 60, the relative errors of the age estimates 
are 20--25 per cent. For SNRs around 20, the values can be as large as 45 per cent or more. 
The median value of relative age errors for the whole sample is 34 per cent. The numerous young 
stars in the sample contribute a significant part of this number. If only stars older than 2\,Gyr are considered, 
the number drops to 30 per cent. 
Fig.\,17 shows the distribution of median stellar age in 1.5$^\circ$$\times$1.5$^\circ$ patches 
on the sky in Galactic coordinate system. As expected, the figure presents a clear positive 
age gradient with increasing Galactic latitudes. Median stellar ages in the disk of $|b|<10^\circ$ 
are younger than 2\,Gyr, while at $|b|>50^\circ$, the median age becomes older than 7\,Gyr.

The distributions of estimated masses and errors of the sample stars are shown in Fig.\,18.
The masses cover a range of 0.7--3.0\,$M_{\odot}$, peaking at 1.1\,$M_{\odot}$. 
Typical errors are smaller than a few per cent for low-mass stars ($<1.5$$M_{\odot}$), 
and about 10 per cent for more massive stars. The median value of the relative mass errors for the 
whole sample is 8 per cent.

\subsection{Age distributions across the [Fe/H]--[$\alpha$/Fe] plane}
 \begin{figure}
\centering
\includegraphics[width=85mm]{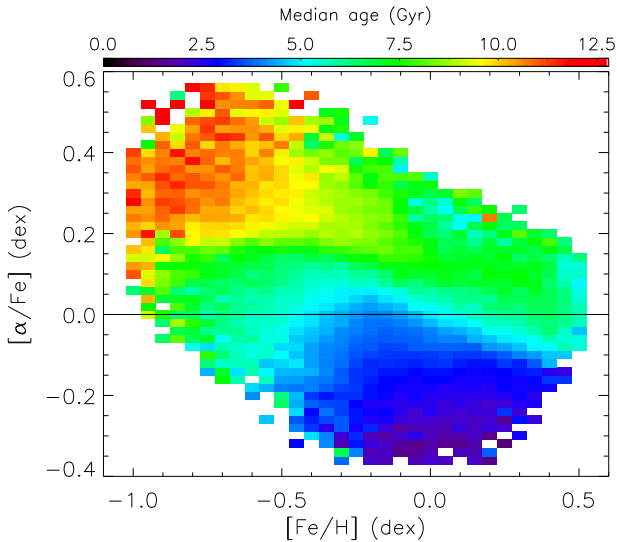}
\caption{Distribution of median ages of stars in mono-abundance bins of 0.05 by 0.02\,dex in the [Fe/H]--[$\alpha$/Fe] 
plane. The horizontal line delineates constant [$\alpha$/Fe] value of 0.0\,dex.}
\label{Fig19}
\end{figure}
 \begin{figure*}
\centering
\includegraphics[width=180mm]{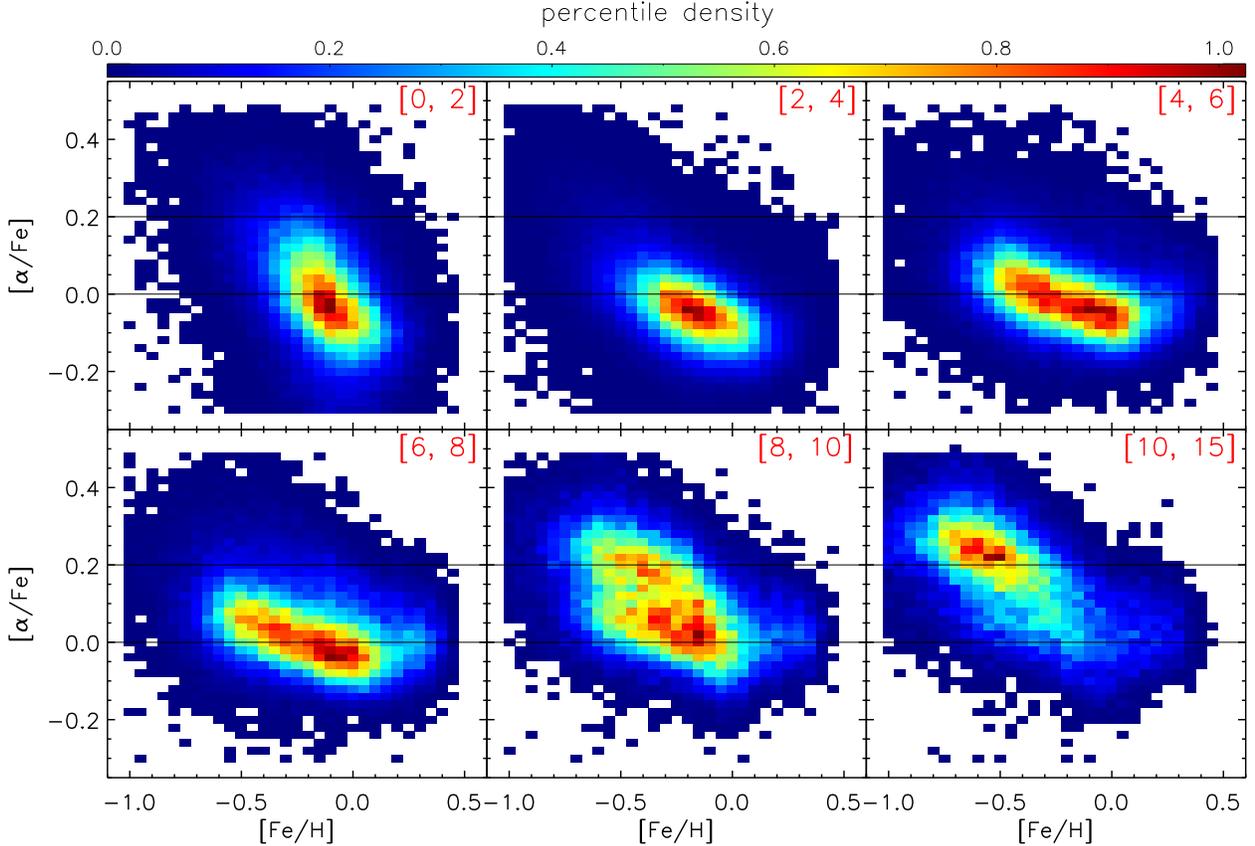}
\caption{Percentile stellar number density in the [Fe/H]--[$\alpha$/Fe] plane for 
stars in different age bins, as labeled in red on the upper-right corner of the figures. 
The horizontal lines delineate constant [$\alpha$/Fe] values of 0.0 and 0.2\,dex, respectively.}
\label{Fig20}
\end{figure*}
Fig.19 plots the median age of sample stars in the individual mono-abundance bins of the [Fe/H]--[$\alpha$/Fe] plane. 
To show potential patterns better, only stars of SNRs higher than 50 are used to generate the figure 
as their atmospheric parameters, especially [$\alpha$/Fe], are estimated with high quality. 
Since precisions of the [$\alpha$/Fe] estimates also vary significantly with 
effective temperatures, as hotter (younger) stars having less precisions due to weaker spectral features 
of [$\alpha$/Fe] indicators, we show results for stars of $T_{\rm eff}<6500$\,K only in the figure. 
After the $T_{\rm eff}$ cut, it is found that there is still a few per cent of young ($\lesssim$4\,Gyr) stars whose 
[$\alpha$/Fe] are artificially overestimated significantly ($>0.2$\,dex), which may cause fake features in the [Fe/H]--[$\alpha$/Fe] 
patterns. Many of those stars are found to have weird spectra in the wavelength range used for the 
[$\alpha$/Fe] estimation (4400--4600 and 5000-5300{\AA}), mainly due to artificial origin 
(e.g. contaminated by nearby bright stars, remains of cosmic ray removal etc.), but some are also
due to intrinsic origin (e.g. composite spectra). 
An effort to identify those weird spectra automatically is still in progress. As a remedy, 
here and below we replace those [$\alpha$/Fe] by [$\alpha$/Fe]$_1$ provided in the value-added catalog, 
which is estimated using spectral wavelength range of 3910--3980, 4400--4600 and 5000-5300{\AA}. 
Specifically, for stars with $T_{\rm eff}$ higher than 5800\,K, if the [$\alpha$/Fe] has a value 
larger than [$\alpha$/Fe]$_1$ by 0.2\,dex, then the [$\alpha$/Fe]$_1$ is adopted. 
The usage of [$\alpha$/Fe]$_1$ effectively reduces the number of young stars with significantly 
overestimated [$\alpha$/Fe].

Fig.\,19 shows clear patterns in the distribution of median stellar ages across the [Fe/H]--[$\alpha$/Fe] plane. 
Generally, more metal-poor and $\alpha$-enhanced stars have 
older ages than metal-rich and $\alpha$-poor ones,  
consistent with previous findings of high resolution spectroscopy of  
solar-neighborhood stars \citep[e.g.][]{Haywood2013, Bergemann2014}. 
The figure further reveals several interesting features. 
Firstly, the most metal-poor (${\rm [Fe/H]}<-0.5$\,dex) and $\alpha$-enhanced (${\rm [\alpha/Fe]}>0.2$\,dex) 
stars are dominated by stars older than 10\,Gyr. Secondly, stars of intermediate-to-old ages (5--8\,Gyr)
show a contiguous distribution across the whole metallicity range from 
$-1.0$ to 0.5\,dex, and exhibit a clear and sharp demarcation from younger and 
more $\alpha$-poor ([$\alpha$/Fe]$\lesssim0.0$) stars. Thirdly, on the 
relatively $\alpha$-poor ([$\alpha$/Fe]$\lesssim0.0$) part of the distribution, 
stellar ages exhibit a gradient with [Fe/H] -- the median ages decrease from $\sim$7--8\,Gyr 
at [Fe/H] of $-0.8$\,dex to 1--2\,Gyr at super-solar metallicities.

Fig.\,20 plots the stellar number density distribution in the [Fe/H] -- [$\alpha$/Fe] plane 
for stars in different age bins. Here a lower SNR cut of 50 is adopted. 
Rather than imposing a temperature cut of 6500\,K as done for Fig.\,19, 
here stars of $T_{\rm eff}<7500$\,K are adopted. This is because the hot ($T_{\rm eff}>6500$\,K) 
stars are mainly distributed in the 0--2\,Gyr bin, and thus do not have an impact on the [Fe/H] -- [$\alpha$/Fe] 
patterns for older stellar populations. 
The figure shows that for all individual age bins, stars exhibit wide distributions in the [Fe/H] -- [$\alpha$/Fe] plane, 
implying that in a given mono-abundance bin of [Fe/H] and [$\alpha$/Fe], stars could have an 
extensive age distribution, especially for bins of intermediate abundances (e.g. $-0.5\lesssim{\rm [Fe/H]}\lesssim0$, 
0$\lesssim$[$\alpha$/Fe]$\lesssim$0.1\,dex). Nevertheless, the figure demonstrates a clear temporal 
evolution trend of [Fe/H] -- [$\alpha$/Fe] sequences. Stars in a relatively young ($<8$\,Gyr) age bin 
distribute along a single sequence with relatively low [$\alpha$/Fe] 
($\lesssim0.1$\,dex), while in the age bin of 10--14\,Gyr, stars distribute mainly along a sequence of high 
[$\alpha$/Fe] ($\gtrsim0.1$\,dex), but with a weak extension to low [$\alpha$/Fe] values (0.0\,dex) 
at solar metallicity. Both the low-$\alpha$ and high-$\alpha$ sequences are presented 
in the age bin of 8--10\,Gyr. As the age increases from 0--2\,Gyr to 8--10\,Gyr, [$\alpha$/Fe] values 
of the lower-$\alpha$ sequence at solar metallicity increase from about $-0.1$\,dex to about 0.0\,dex. 
Note that the [Fe/H] -- [$\alpha$/Fe] sequence of the youngest stars (0--2\,Gyr) exhibits a steeper slope, 
which is probably due to problematic [$\alpha$/Fe] estimates for such young (hot) stars.  
From 8--10\,Gyr to 10--14\,Gyr, it seems that the high-$\alpha$ sequence extends to  
lower metallicity (by 0.1--0.2\,dex) and slightly ($\lesssim$0.05\,dex) higher [$\alpha$/Fe] values.  
Such a double-sequence feature is consistent with the widely suggested thin and thick disk sequences 
\citep[e.g.][]{Fuhrmann1998, Bensby2003, Lee2011, Haywood2013, Hayden2015}. 
Our results thus suggest that the Galactic thin disk became a prominent structure at 8--10\,Gyr ago, 
while the Galactic thick disk formed at earlier epoch and was almost quenched at about 8\,Gyr ago. 

\subsection{The age--[$\alpha$/Fe] and age--[$\alpha$/H] correlation}
\begin{figure*}
\centering
\includegraphics[width=170mm]{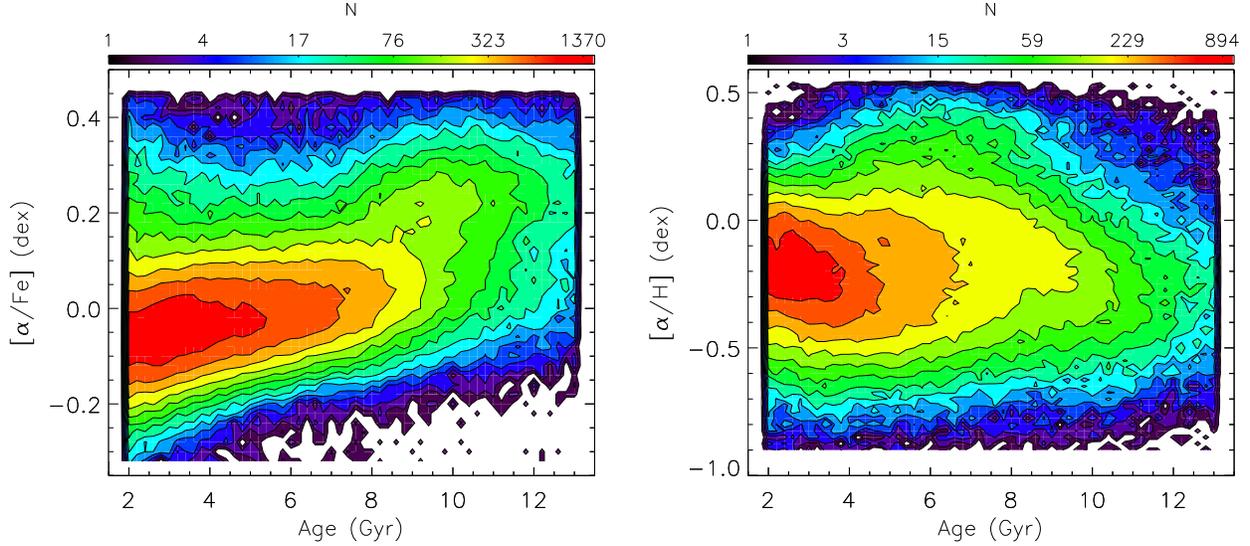}
\caption{Color-coded stellar number density distribution in the age--[$\alpha$/Fe] (left) and age--[$\alpha$/H] 
(right) planes. The densities are shown in logarithmic (base 10) scale.}
\label{Fig21}
\end{figure*}
Fig.\,21 plots the density distribution of the sample stars 
in the age--[$\alpha$/Fe] and age--[$\alpha$/H] planes. Here the [$\alpha$/H] 
is converted from [$\alpha$/Fe] and [Fe/H]. Only stars of ${\rm SNRs}>50$ 
and $T_{\rm eff}<7500$\,K are used to ensure reliable [$\alpha$/Fe] estimates. 
Moreover, as has been discussed above, the current [$\alpha$/Fe] estimates for 
hot (young) stars are likely problematic, so that we further discard stars younger than 2\,Gyr. 
Note that despite these measures, some young (e.g. $<4$\,Gyr) stars with 
problematic [$\alpha$/Fe] estimates (e.g. $>0.2$\,dex) still remain in the figure. 
The figure shows two sequences in the age--[$\alpha$/Fe] plane. 
Stars younger than 8\,Gyr belong to a sequence of lower [$\alpha$/Fe] values, 
and the [$\alpha$/Fe] slowly increases with age in an approximately linear manner 
with a slope of $\lesssim$0.02\,dex/Gyr. At the older end, the low-$\alpha$ 
sequence extends to an age older than 10\,Gyr. There is also a sequence 
with higher [$\alpha$/Fe], which has an almost constant [$\alpha$/Fe] value of about 
0.25($\pm$0.05)\,dex for stars older than 10\,Gyr. At the younger end, the high-$\alpha$ 
sequence extends to $\sim$8\,Gyr, when it connects with the low-$\alpha$ sequence, 
consistent with results from Fig.\,20. The presence of two age--[$\alpha$/Fe] sequences 
either suggests the existence of two distinct phases of formation history of the Galactic disk \citep[e.g.][]{Haywood2013, Xiang+2015b} 
or is a natural consequence of a continuous disk formation process \citep[e.g.][]{Schonrich2009}. 
Whatever processes have caused the multiple age--[$\alpha$/Fe] relations, 
it seems that 8--10\,Gyr is a special epoch in the disk formation history. 

The age--[$\alpha$/H] plane exhibits a significant lack of old ($>8$\,Gyr), $\alpha$-rich ($>0.0$)
stars, leading to a negative age--[$\alpha$/Fe] sequence at early time.  At the younger end, 
it seems that the sequence extends to $\sim6$\,Gyr, when the [$\alpha$/H] reaches a maximum value of 
0.3--0.4\,dex. At any given age younger than 8\,Gyr, the [$\alpha$/H] exhibits a wide distribution. 
Nevertheless, it seems that stars younger than 5\,Gyr follow an overall negative age--[$\alpha$/H] 
sequence, rather than a flat one. This younger sequence has a median [$\alpha$/H] of about 
$-0.3$\,dex at an age of 5\,Gyr, and reaches a median [$\alpha$/H] value of $-0.2$\,dex at 2\,Gyr. 
At intermediate age range 5--8\,Gyr, overlaps of the two sequences seem to have smoothed 
the negative age--[$\alpha$/H] trends. At the high-[$\alpha$/H] end, the contours show 
positive slopes, probably a natural consequence of the overlapping of the two sequences. 

\subsection{The age -- metallicity  relation}
 \begin{figure}
\centering
\includegraphics[width=85mm]{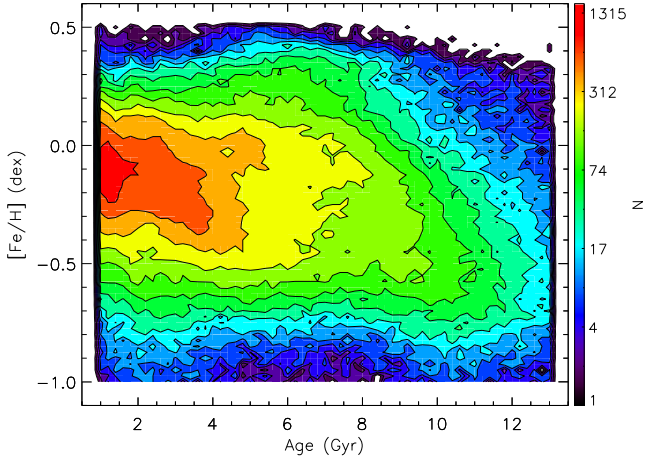}
\caption{Color-coded stellar number density distribution in the age--[Fe/H] plane. 
The densities are shown in logarithmic (base 10) scale.}
\label{Fig22}
\end{figure}
Fig.\,22 plots the density distribution of the sample stars in the age--[Fe/H] plane. 
To ensure small uncertainties in age and [Fe/H] estimates thus to better illustrate 
systematic trends, only stars of ${\rm SNRs}>50$ and $T_{\rm eff}<8000$\,K are shown. 
Stars younger than 1\,Gyr are discarded for completeness (in [Fe/H]) reasons.
That is, as the temperature of a MSTO star depends sensitively on both age and metallicity, 
a $T_{\rm eff}$ cut of 8000\,K discards more metal-poor young ($<1$\,Gyr) stars than metal-rich 
ones, thus leads to undesired trends in the age range 0--1\,Gyr.
The figure shows a wide range of [Fe/H] at all ages younger than 8\,Gyr.  
At the older end ($>8$\,Gyr), there is an obvious lack of metal-rich stars, yielding a relatively 
tight age--[Fe/H] correlation. The patterns are in good agreement with previous findings for 
stars in the solar-neighbourhood \citep[e.g.][]{Haywood2013, Bergemann2014}. 
The relatively tight age--[Fe/H] correlation for old disk stars implies that 
at any given time, the interstellar medium forming the stars was relatively well-mixed.   
On the other hand, the broad range of [Fe/H] values for young disk stars at a given age 
suggests a more complicated chemical enrichment history. 
As the sample stars cover a large volume, one possible cause of the broad [Fe/H] distribution 
is the existence of both radial and vertical [Fe/H] gradients for mono-age stellar populations \citep{Xiang+2015c}.
However, it is also found that even in a limited volume, for instance, the solar neighborhood, 
the age--[Fe/H] relation for young ($<8$\,Gyr) stars exhibits still a broad distribution.  
The inevitable presence of mixing of stars born at different positions (thus with different values of [Fe/H]) caused by 
stellar radial migration \citep[e.g.][]{Sellwood2002, Roskar2008, Schonrich2009b, Loebman2011} 
has certainly played a role for such [Fe/H] broadening. Whereas for very young (e.g. $\sim1$\,Gyr) stars, 
the broad [Fe/H] distribution is probably largely caused by sustained star formation process via accreting 
metal-poor gas from outside the disc, as the timescale is too short for radial migration to 
make a great impact.

In addition to the above qualitative patterns in agreement with the previous findings, 
the current large sample also reveals several interesting features. Firstly, rather than 
a `flat' age--[Fe/H] relation, as suggested by the previous studies \citep[e.g.][]{Bergemann2014}, 
young ($<5$\,Gyr) disk stars seems to exhibit a negative overall trend of [Fe/H] with age, similar to that found for [$\alpha$/H].  
Few studies exist on such a possible negative age--[Fe/H] trend for the young disk stars 
due to the limited size of stellar sample available previously. A further, more careful analysis 
shows that the slopes of the age--[Fe/H] relations of young stars vary with Galactocentric radius. 
In the outer disk, the negative age--[Fe/H] relation becomes steeper. 
The observed age--[Fe/H] relation of the Galactic disk is thus, similar to the age--[$\alpha$/H] 
relation, composed of at least two negative sequences, one for old ($\gtrsim8$\,Gyr) stars and 
another for young ($\lesssim$5\,Gyr) stars. At the intermediate age range 5--8\,Gyr, 
mixing of stars that follow the two separate sequences makes the trend less distinct. 
This two distinct sequences of age--[Fe/H] relation, if confirmed, may provide important constrains 
on the chemical enrichment history of the Galactic disk -- it is possible that they  
result from two different global chemical enrichment processes of the Galactic disk. 
Interestingly, utilizing about 20,000 subgiant stars selected from the LAMOST DR2, \citet{LiuC2015} 
find evidence of a `narrow stripe' of stars alongside with the `main stripe' stars 
in the age--[Fe/H] plane. They interpret those `narrow stripe' stars as migrators.  
However, we note that their sample is less complete than ours in the age--[Fe/H] plane  
as their sample lacks both metal-rich and metal-poor young ($<4$\,Gyr) stars (see their Fig.\,7). 
Secondly, there is a considerable fraction of intermediate-aged (5--8\,Gyr) stars of 
${\rm [Fe/H]}>0.3$\,dex. As a result, the density contours at the metal-rich end 
show positive slopes. One possible origin of these positive trends 
is that those metal-rich stars are migrators from the inner disk -- the older 
the stars the further from the inside as they have longer time to reach their current positions \citep{Loebman2011}. 
Alternatively, a natural explanation of those positive trends is the mixing of stars 
from the two sequences of age--[Fe/H] relations -- the intermediate-aged stars 
of ${\rm [Fe/H]}>0.3$\,dex belong to the older sequence, whereas the young, 
metal-rich stars are mainly composed of stars following the younger sequence.
In addition, there are a considerable number of young ($<4$\,Gyr) metal-poor stars 
in our sample, some of them can be as metal-poor as ($\lesssim-0.6$\,dex) 
the oldest ($>10$\,Gyr) ones. The origin of those young yet metal-poor stars, 
as well as that for the young [$\alpha$/H]-poor stars in Fig.\,21, needs to be further studied. 
Finally, we note that the distribution also shows a number of substructures/over-densities 
whose genuineness and origins remain to be further studied. 
 
\subsection{Distribution of stellar ages in the $R$ -- $Z$ plane}
\begin{figure}
\centering
\includegraphics[width=85mm]{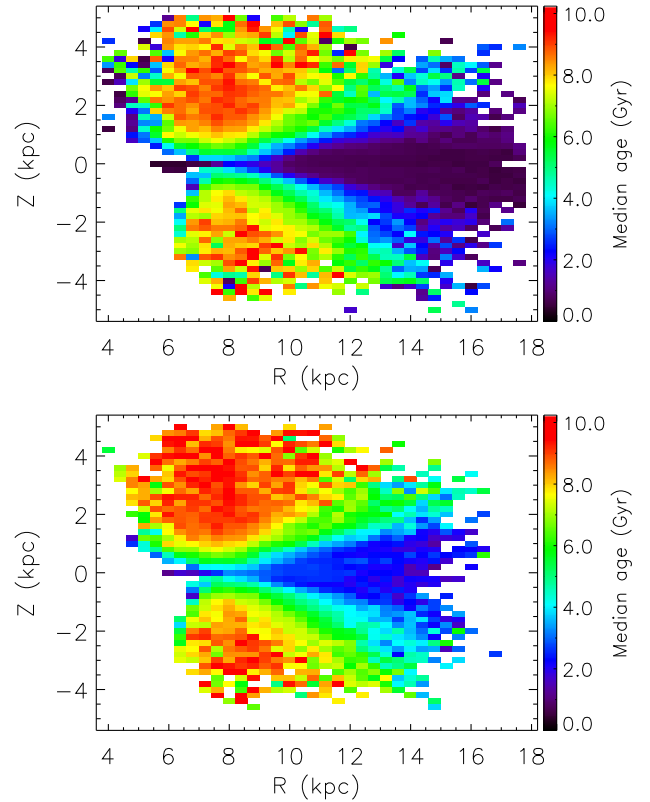}
\caption{Color-coded distributions of the median age for stars 
in different spatial bins of the $R-Z$ plane. The adopted bin size is 0.4\,kpc in the $R$ direction 
and 0.2\,kpc in the $Z$ direction. The upper panel shows results of the whole sample 
stars, while the lower panel is for sample stars of $T_{\rm eff}<7000$\,K.}
\label{Fig23}
\end{figure}
Fig.\,23 plots the median age of stars at different positions across the $R$--$Z$ plane 
of the Galactic disk. Here $R$ is the projected Galactocentric distance in the disk mid-plane, 
and $Z$ the height above the disk mid-plane. The top panel presents results from all of the sample stars. 
Generally, the data exhibit negative age gradients in the radial and 
positive age gradients in the vertical direction. At small heights, the outer disk of 
$R\gtrsim9$\,kpc is dominated by young ($\lesssim$2\,Gyr) stars, which reach 
larger heights above the disk plane at the farther disk, which exhibits 
a strong flare in median stellar age. The inner disk ($R\lesssim9$\,kpc) exhibits 
a positive vertical age gradient for $|Z|\lesssim1$\,kpc, 
while at larger heights above the disk plane, old ($\gtrsim$10\,Gyr) stars dominate 
the population with no significant vertical gradients. However, for many bins 
near the boundary of the $R-Z$ plane covered by the sample stars, the stellar 
populations are dominated by unexpected young to intermediate-aged stars. 
At large height, e.g. $Z>2$\,kpc, those unexpected features are likely caused by 
blue stragglers whose ages have been artificially underestimated, as will be discussed in \S{7}. 
Those stars are usually hot and bright thus can be detected at large distances. 
To reduce the contaminations of blue stragglers, in the bottom panel of Fig.\,23, 
we present the age distribution after excluding stars of $T_{\rm eff}>7000$\,K. 
The result shows much more clean patterns, and the unexpected young populations 
at large heights in the inner disk now largely disappear. On the other hand, 
since intrinsically young stars are also discarded by the temperature cut, the 
outer disk exhibits systematically older ages compared to those shown in the top panel. 
Nevertheless, the overall structures and patterns remain unaffected. 

A radial age gradient of the geometrically thick disk was also presented 
by \citet{Martig2016b} using giants from the APOGEE survey. 
A flaring young stellar disk in the outer part has been observed previously via star counts
\citep[e.g.][]{Derriere2001, Lopez-Corredoira2002, Lopez-Corredoira2014} and  
well reproduced by simulations \citep[e.g.][]{Narayan2002, Rahimi2014, Minchev2015} 
as a suggested consequence of weaker restoring force at the outer Galactocentric radii. 
Nevertheless, Fig.\,23 demonstrates the first explicit picture of disk flare in stellar age, 
which will provide further constrains on disk flare models. 

Note however that our results are no doubt affected by some selection effects since 
the sample is a magnitude limited one. Younger stars tend to be brighter thus probe 
larger volume than older, fainter stars. The unexpected young stellar ages near the 
boundary of the $R$-$Z$ plane at small heights of the inner disk are likely due to 
such selection effects.  Age distribution in the outer disk is probably also suffering 
severe biases due to selection effects. A detailed and quantitative 
study of the selection effects of our sample stars is beyond the scope of this paper 
and will be presented elsewhere. 

\section{Unresolved binaries and blue stragglers}
 \begin{figure*}
\centering
\includegraphics[width=160mm]{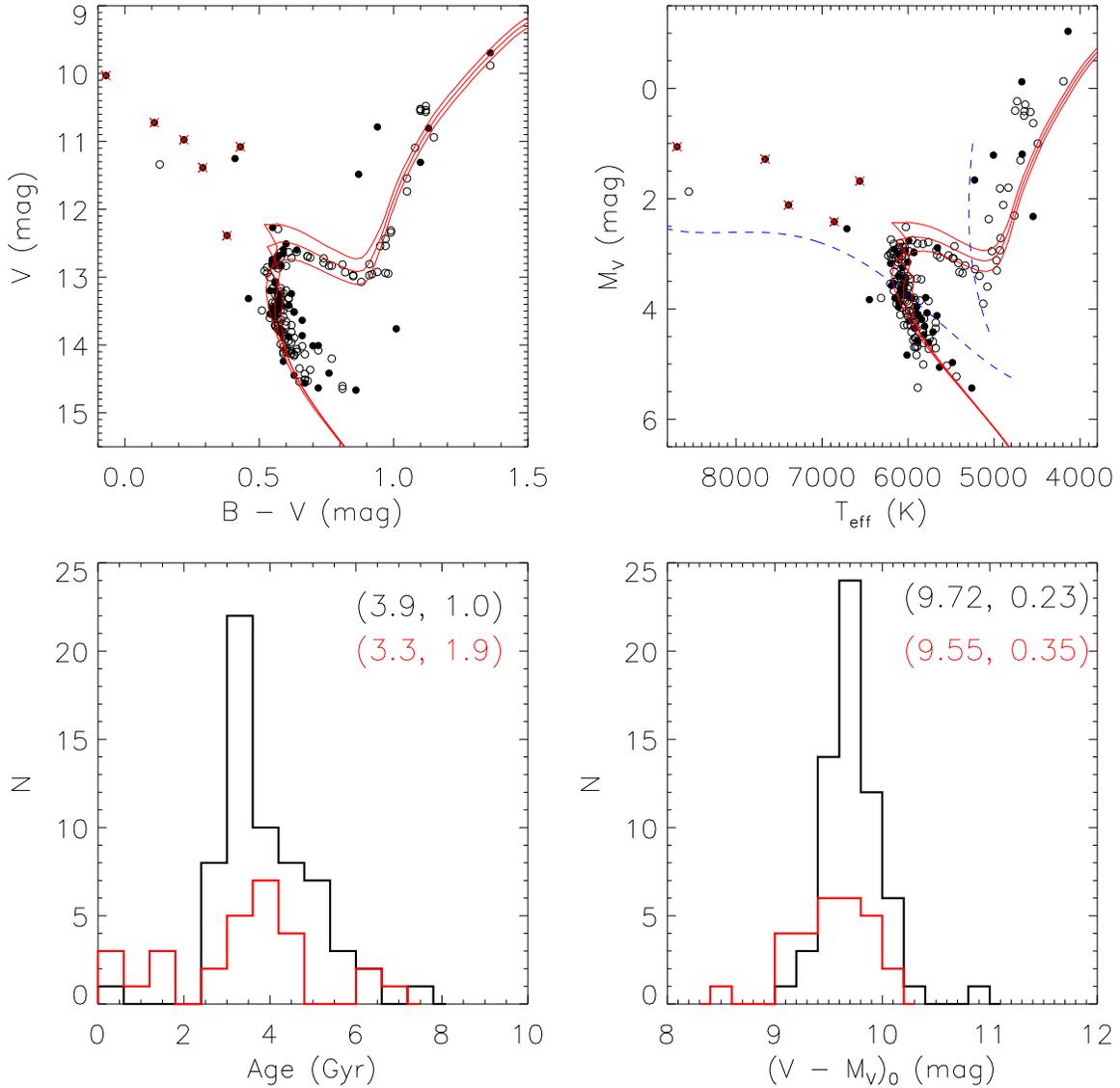}
\caption{$Upper$: Distributions of M67 member stars from \citet{Geller2015} that  
have LAMOST observations in the ($B-V$, $V$) color-magnitude diagram (left) and the 
$T_{\rm eff}$ -- ${\rm M}_V$ diagram (right). Single stars and binaries are shown 
by empty and filled circles, respectively. Crosses in red represent blue stragglers. 
The isochrones are for ages 3.5, 4.0 and 4.5\,Gyr, respectively, with 
[Fe/H] = 0 and [$\alpha$/Fe]=0\,dex. A distance modulus of 9.7\,mag and a color excess $E(B-V)$ of 0.03\,mag 
are adopted to place the isochrones in the ($B-V$, $V$) diagram. The dashed 
lines in the $T_{\rm eff}$--${\rm M}_V$ diagram denote the MSTO-SG 
star selection criteria for solar metallicity. $Lower$: Distributions of ages (left) 
and distance moduli (right) for single (black) and binary (red) members. 
The mean and standard deviation of the age and modulus for both single 
and binary members are marked in the plots.}
\label{Fig24}
\end{figure*}
About 40 per cent of the Galactic field stars are found in binary systems \citep[e.g.][]{Gao2014, Yuan2015c}. 
The distance modulus of an unresolved binary is usually underestimated if treated as a single star, 
and the amount of underestimation depends on the mass ratio of the binary components, 
reaching a maximum of 0.75\,mag in case of equal mass. This may introduce potential bias 
to the MSTO-SG sample. However, because binaries of 
high mass ratios contribute only a small fraction of the whole binary population \citep[e.g.][]{Duchene2013}, 
we expect that the fraction of stars whose distances have been significantly overestimated is small in our sample. 
Stellar parameters of a binary could also be wrongly estimated by the current stellar parameter pipelines. 
Fortunately, exercises show that $T_{\rm eff}$, ${\rm M}_V$ and [Fe/H] derived with 
LSP3 from a binary spectrum are only marginally different to those derived from the spectrum 
of the main component star, with typical differences of only a few tens Kelvin in 
$T_{\rm eff}$, $<0.1$\,dex in [Fe/H] and $0.1$\,mag 
in ${\rm M}_V$. This is consistent with the finding of \citet{Schlesinger2010}, 
who have analyzed the effects of binaries on stellar parameter determinations with the SDSS/SEGUE spectra. 

As an examination, the top right panel of Fig.\,24 plots distributions of single and binary member 
stars of M67 from \citet{Geller2015} in the $T_{\rm eff}$ -- ${\rm M}_V$ diagram, with $T_{\rm eff}$ 
and M$_V$ derived from LAMOST spectra with LSP3. 
The single and binary member stars of \citet{Geller2015} are classified using precise radial velocity measurements. 
There are 142 unique single stars and 58 unique binaries that have LAMOST spectra with SNRs 
higher than 20. Fig.\,24 shows that most of the 
binaries follow the same locus with single stars in both the color-magnitude and $T_{\rm eff}$--${\rm M}_V$ diagrams. 
Among those members, 62 single and 34 binary stars pass our selection criteria of MSTO-SG stars. 
Most single and binary MSTO-SG stars have similar age distributions, with a mean age of $\sim$4.0\,Gyr 
and a standard deviation of $\sim$1.0\,Gyr. There are a few outliers in the age distribution 
of binary members. Among them, the younger ones are blue stragglers, while the older ones 
are likely contaminations of main sequence binary stars with large parameter uncertainties. 
The distributions of distance moduli of the single and binary populations exhibit some discrepancies, 
in the sense that binaries yield a mean modulus 0.17\,mag smaller than single stars due to 
their brighter apparent magnitudes.
The difference corresponds to a $\sim$10 per cent underestimation of their distances.   

Blue straggler stars (BSS) are generally believed to be products of  coalescence or mass exchange 
in binary evolution \citep[e.g.][]{Chen2008a, Chen2008b}, and as a result, they are more luminous 
and bluer than MSTO stars of the same age. The ages of those stars may have been  
artificially underestimated with our current method. 
An accurate determination of the fraction of BSS with respect to the whole stellar population 
has not been carried out, although we expect this number to be considerable considering that 
there are large number of such stars in our sample. As for M67 shown in Fig.\,24, 6--8 out of the 
98 MSTO-SG stars ($\sim$7 per cent) are BSS according to the identification of \citet{Geller2015}. 
For our MSTO-SG star sample, exercises show that at $|Z|>1.5$\,kpc, about 4 per cent stars 
have age estimates younger than 2\,Gyr. 
Since young stars are not expected to occur at such a large height away from the disk midplane, 
they are likely BSS stars whose ages have been artificially underestimated.  
Contamination rate of BSS to truly young stars having comparable ages as estimated 
for the BSS should be even higher and needs to be further studied. As a rough estimate, we can 
simply assume that for sample stars of 2--12\,Gyr, 4 per cent are actually BSS 
whose ages have been wrongly underestimated to be younger than $2$\,Gyr. If we further assume that 
the star formation rate of the Milky Way disk is flat, then the total contamination of BSS to the 
young sample stars ($<2$\,Gyr) is about 20 per cent.

\section{Summary}
A sample of 0.93 million disk MSTO and subgiant stars  
are defined using the value-added catalog of the LAMOST Galactic Spectroscopic Surveys. 
Stellar masses and ages of the sample
stars are estimated with a Bayesian algorithm based on stellar isochrones.
Uncertainties of the resultant parameters depend on the SNR, and have a typical (median)
value of $0.08$\,${\rm M}_\odot$ in mass and 34 per cent in age, and half of the sample 
stars older than 2\,Gyr have age uncertainties of only 20--30 per cent. 
Robustness of the results are validated via examinations with extensive 
datasets, including a mock dataset, the LAMOST-TGAS common stars, LAMOST-$Kepler$ 
common stars that have asteroseismic parameters, member stars of open clusters 
as well as duplicate observations of the sample stars. In addition to the random errors, 
there is probably also a systematic uncertainty of about 1--2\,Gyr in the age estimates 
as a consequence of inadequate model assumptions of stellar isochrones as well as 
inadequacy in the analysis method. For the young stellar populations,
contaminations from blue straggler stars are non-negligible.

Interesting patterns are seen in the distribution of median age across the [Fe/H]--[$\alpha$/Fe] plane.  
As expected, metal-poor, $\alpha$-enhanced stars are generally older than metal-rich, 
$\alpha$-poor stars. The most metal-poor (${\rm [Fe/H]}<-0.5$\,dex) and $\alpha$-enhanced 
(${\rm [\alpha/Fe]}>0.2$\,dex) stars have median ages older than 10\,Gyr. Stars 
of intermediate-to-old ages (5--8\,Gyr) exhibit a contiguous distribution across the whole 
metallicity range from $-1.0$ to 0.5\,dex and have a clear demarcation from younger, 
more $\alpha$-poor ([$\alpha$/Fe]$\lesssim0.0$\,dex) stars. The latter shows an age 
gradient with [Fe/H]. 
Stellar density distribution in the [Fe/H]--[$\alpha$/Fe] plane exhibits
both thin and thick disk sequences for stars of 8--10\,Gyr, while only the thin (thick) disk 
sequence is present for younger (older) stars, implying that the thin disk became 
prominent 8--10\,Gyr ago, while the thick disk was formed at an earlier epoch and 
was almost quenched 8\,Gyr ago.

Good correlations between age and [$\alpha$/Fe] or [$\alpha$/H] (and [Fe/H]) are found. 
There are two correlation sequences in the age--[$\alpha$/Fe] plane. 
The lower-$\alpha$ sequence contains stars with ages ranging from younger than 1 to 
older than 10\,Gyr, and the [$\alpha$/Fe] values slowly increase with age in an approximately linear 
manner with a slope of $\lesssim$0.02\,dex/Gyr. The higher-$\alpha$ sequence 
is composed of relatively old ($>8$\,Gyr) stars, and the [$\alpha$/Fe] values are 
almost constant ($\sim$0.25\,dex) for stars older than 10\,Gyr, and then decrease 
with age in the range 8--10\,Gyr.
The sample stars exhibit also two sequences in the age--[Fe/H] and age--[$\alpha$/H] planes. 
There is a lack of metal-rich stars older than 8\,Gyr, which forms a negative age--[Fe/H] sequence. 
This older sequence seems to reach $\sim$5--6\,Gyr at the younger end, with [Fe/H] 
increasing from $\sim-0.6$\,dex at 12\,Gyr to $\sim$0.3\,dex at 6\,Gyr.
Instead of a `flat' distribution, the huge sample 
reveals also a significant negative age--[Fe/H] sequence for stars younger 
than $\sim$5\,Gyr. At intermediate ages of about 5--8\,Gyr, mixing of stars 
from the two sequences makes the negative age--[Fe/H] correlations less obvious. 
Similar trends are seen in the age--[$\alpha$/H] plane. Moreover, at the lower-[$\alpha$/H] 
side, the younger sequence seems to have [$\alpha$/H] values lower than those of the 
older sequence. 

The sample stars exhibit interesting age structures across the disk of $4<R<18$\,kpc.
The median stellar age increases with $Z$ and decreases with $R$, yielding a positive 
age gradient in the vertical and a negative gradient in the radial direction of the disk. 
At the outer disk of $R\gtrsim9$\,kpc, the stellar age shows a strong flaring structure, 
which is expected to provide further constrains on disk flare models. 

\vspace{7mm} \noindent {\bf Acknowledgments}{
It is a pleasure to thank the anonymous referee for valuable comments and suggestions.
This work is supported by Joint Funds of the National Natural Science Foundation of China 
(Grant No. U1531244) and National Key Basic Research Program of China 2014CB845700.  
Z.-Y. Huo and M.-S. Xiang acknowledge supports from the National Natural Science 
Foundation of China (NSFC, Grant No. 11403038). H.-B. Yuan acknowledges supports 
from NSFC grant No. 11443006, No. 11603002 and Beijing Normal University grant No. 310232102. 
The LAMOST FELLOWSHIP is supported by Special Funding for Advanced Users, 
budgeted and administrated by Center for Astronomical Mega-Science, 
Chinese Academy of Sciences (CAMS). 

Guoshoujing Telescope (the Large Sky Area Multi-Object Fiber
Spectroscopic Telescope LAMOST) is a National Major Scientific
Project built by the Chinese Academy of Sciences. Funding for
the project has been provided by the National Development and
Reform Commission. LAMOST is operated and managed by the National
Astronomical Observatories, Chinese Academy of Sciences.}

\bibliographystyle{mn2e}
\bibliography{MSTO.bib}


\newpage
\appendix
\section {}
\begin{table*}
\caption{Trajectories of MSTO in the $T_{\rm eff}$ -- ${\rm M}_V$ plane, 
${\rm M}_V^{\rm TO}$ = $a_0$ + $a_1\times$$T_{\rm eff}$ + $a_2\times$$T_{\rm eff}^2$ + $a_3\times$$T_{\rm eff}^3$ + $a_4\times$$T_{\rm eff}^4$, 
as well as the adopted minimum temperature of MSTO of isochrones, $T_{\rm eff}^{\rm MINISO}$. }
\begin{tabular}{rrrrrrrr}
\hline
[Fe/H] &  [$\alpha$/Fe] & $a_0$ & $a_1$ & $a_2$ & $a_3$ & $a_4$ & $T_{\rm eff}^{\rm MINISO}$ (K) \\
\hline
     $-1.0$ &       0.30 &       32.9210 &   $-0.00828$ &   5.10153e$-07$ &   2.79169e$-11$ &  $-$2.79516e$-15$  & 5678 \\
    $-0.9$ &       0.28 &       24.8208 &   $-0.00376$ &  $-$4.27601e$-07$ &   1.13254e$-10$ &  $-$5.67469e$-15$ & 5637  \\
    $-0.8$ &       0.26 &       15.5995 &    0.00139 &  $-$1.50012e$-06$ &   2.11171e$-10$ &  $-$8.99138e$-15$  & 5594\\
    $-0.7$ &       0.24 &       6.4951 &    0.00655 &  $-$2.58761e$-06$ &   3.11700e$-10$ &  $-$1.24379e$-14$  & 5546 \\
    $-0.6$ &       0.22 &      $-3.6820$ &     0.01233 &  $-$3.80997e$-06$ &   4.25071e$-10$ &  $-$1.63391e$-14$ & 5497 \\
    $-0.5$ &       0.20 &      $-13.6972$ &     0.01810 &  $-$5.04614e$-06$ &   5.41152e$-10$ &  $-$2.03809e$-14$ & 5450 \\ 
    $-0.4$ &       0.16 &      $-24.1542$ &     0.02418 &  $-$6.35876e$-06$ &   6.65373e$-10$ &  $-$2.47385e$-14$ & 5416 \\
    $-0.3$ &       0.12 &      $-35.6453$ &     0.03088 &  $-$7.81131e$-06$ &   8.03316e$-10$ &  $-$2.95945e$-14$ & 5377 \\
    $-0.2$ &       0.08 &      $-46.9116$ &     0.03755 &  $-$9.27544e$-06$ &   9.44030e$-10$ &  $-$3.46030e$-14$ & 5335 \\
    $-0.1$ &       0.04 &      $-59.0560$ &     0.04477 &  $-$1.08696e$-05$ &   1.09792e$-09$ &  $-$4.01032e$-14$ & 5268 \\
      0.0 &       0.00 &      $-71.1208$ &     0.05205 &  $-$1.24941e$-05$ &   1.25640e$-09$ &  $-$4.58219e$-14$ & 5253 \\
     0.1 &       0.00 &      $-83.8769$ &     0.05979 &  $-$1.42343e$-05$ &   1.42708e$-09$ &  $-$5.20100e$-14$  & 5179 \\
     0.2 &       0.00 &      $-96.9268$ &     0.06781 &  $-$1.60510e$-05$ &   1.60669e$-09$ &  $-$5.85678e$-14$  & 5175 \\
     0.3 &       0.00 &      $-110.4560$ &     0.07619 &  $-$1.79643e$-05$ &   1.79699e$-09$ &  $-$6.55526e$-14$ & 5130  \\
     0.4 &       0.00 &      $-124.7290$ &     0.08510 &  $-$2.00142e$-05$ &   2.00203e$-09$ &  $-$7.31135e$-14$ & 5096 \\
     0.5 &       0.00 &      $-139.8100$ &     0.09455 &  $-$2.21929e$-05$ &   2.22051e$-09$ &  $-$8.11890e$-14$ & 5287 \\
\hline
\end{tabular}
\end{table*}

\begin{table*}
\caption{Trajectories of base RGB in the $T_{\rm eff}$ -- ${\rm M}_V$ plane, 
$T_{\rm eff}^{\rm bRGB}$ = $b_0$ + $b_1\times{\rm M}_V$ + $b_2\times{\rm M}_V^2$ + $a_3\times{\rm M}_V^3$. }
\begin{tabular}{rrrrrr}
\hline
[Fe/H] &  [$\alpha$/Fe] & $b_0$ & $b_1$ & $b_2$ & $b_3$  \\
\hline
     $-1.0$ &     0.30 &       5156.16 &       107.612 &      $-5.41575$ &      $-6.10802$ \\
    $-0.9$ &      0.28 &       5126.67 &       111.224 &      $-7.62809$ &      $-5.80089$ \\
    $-0.8$ &      0.26 &       5096.75 &       112.755 &      $-8.75758$ &      $-5.62243$ \\
    $-0.7$ &      0.24 &       5068.06 &       111.355 &      $-8.75119$ &      $-5.55727$ \\
    $-0.6$ &      0.22 &       5039.51 &       109.027 &      $-8.36525$ &      $-5.52200$ \\
    $-0.5$ &      0.20 &       5010.91 &       106.143 &      $-7.72652$ &      $-5.49936$ \\
    $-0.4$ &      0.16 &       4982.45 &       102.590 &      $-6.79953$ &      $-5.48756$ \\
    $-0.3$ &      0.12 &       4954.27 &       98.2728 &      $-5.55812$ &      $-5.48322$ \\
    $-0.2$ &      0.08 &       4925.72 &       94.5977 &      $-4.56288$ &      $-5.42086$ \\
    $-0.1$ &      0.04 &       4896.64 &       92.3138 &      $-4.14167$ &      $-5.25891$ \\
      0.0 &       0.00 &       4868.28 &       89.1947 &      $-3.41834$ &      $-5.08543$ \\
     0.1 &       0.00 &       4840.10 &       86.5965 &      $-2.94804$ &      $-4.83513$ \\
     0.2 &       0.00 &       4812.81 &       83.2880 &      $-2.25062$ &      $-4.55325$ \\
     0.3 &       0.00 &       4786.01 &       80.3599 &      $-1.76849$ &      $-4.18724$  \\
     0.4 &       0.00 &       4760.20 &       77.1021 &      $-1.23815$ &      $-3.75894$  \\
     0.5 &       0.00 &       4735.49 &       73.5323 &     $-0.679470$ &      $-3.26016$  \\
\hline
\end{tabular}
\end{table*}

\end{document}